\documentclass[sigplan,screen,prologue,table]{acmart}
\usepackage{pifont}
\usepackage{xspace}
\usepackage{multirow} 
\usepackage{nicematrix}
\usepackage{booktabs}
\usepackage{subfig}
\newcommand{\PaperAcronym}{D3Bench\xspace}
\newcommand{\quotes}[1]{``#1''}
\AtBeginDocument{%
  \providecommand\BibTeX{{%
    \normalfont B\kern-0.5em{\scshape i\kern-0.25em b}\kern-0.8em\TeX}}}

\begin{document}

\title{Open-Source Drift Detection Tools in Action: \\Insights from Two Use Cases}

\author{Rieke Müller}
\affiliation{%
  \institution{Software AG}
  \streetaddress{Uhlandstrasse 12}
  \city{Darmstadt}
  \state{Hessen}
  \country{Germany}
  \postcode{64297}
}
\email{Rieke.Mueller@softwareag.com}

\author{Mohamed Abdelaal}
\affiliation{%
  \institution{Software AG}
  \streetaddress{Uhlandstrasse 12}
  \city{Darmstadt}
  \country{Germany}}
\email{Mohamed.Abdelaal@softwareag.com}

\author{Davor Stjelja}
\affiliation{%
  \institution{Granlund Oy}
  \city{Helsinki}
  \country{Finnland}
}
\email{Davor.Stjelja@granlund.fi}

\renewcommand{\shortauthors}{Mueller, et al.}

\begin{abstract}
Data drifts pose a critical challenge in the lifecycle of machine learning (ML) models, affecting their performance and reliability. In response to this challenge, we present a microbenchmark study, called \PaperAcronym{}, which evaluates the efficacy of open-source drift detection tools. \PaperAcronym examines the capabilities of Evidently AI, NannyML, and Alibi-Detect, leveraging real-world data from two smart building use cases. We prioritize assessing the functional suitability of these tools to identify and analyze data drifts. Furthermore, we consider a comprehensive set of non-functional criteria, such as the integrability with ML pipelines, the adaptability to diverse data types, user-friendliness, computational efficiency, and resource demands. Our findings reveal that Evidently AI stands out for its general data drift detection, whereas NannyML excels at pinpointing the precise timing of shifts and evaluating their consequent effects on predictive accuracy.
\end{abstract}

\maketitle

\section{Introduction}\label{sec:intro}
%
Machine learning nowadays enhances our everyday digital interactions, with applications ranging from personalized recommendations to predictive keyboards~\cite{sarker2021machine}. It plays a transformative role in healthcare and transportation, improving patient outcomes and optimizing logistics, respectively. In essence, ML algorithms operate as an invisible but indispensable assistant, streamlining complex tasks, offering insights from vast amounts of data, and making technology more adaptive and responsive to our human needs. The efficacy of ML algorithms is broadly intertwined with the quality of the data used for both training and serving purposes~\cite{rein23}. 

A particular challenge that can compromise both model performance and data integrity is the phenomenon of data drift. In the traditional development of supervised learning algorithms, it is often assumed that the distribution of training data, test data, and production data is the same. Data drifts occur when the statistical properties of the serving data diverge from those of the training data, leading to a decline in model accuracy and, consequently, the reliability of its outputs~\cite{subbaswamy2022unifying}. These differences may be due to a model being used in a new location for which no data was available during training, or due to natural variations that occur over time. The environmental conditions or the patterns according to which the model predicts may also have changed and this can lead to a decrease in the accuracy of the model. In several application areas, the COVID-19 pandemic has shown how data drifts can lead to model failures. For example, during the outbreak of the pandemic, Amazon's supply chain prediction algorithms failed due to the sudden increase in demand for household goods, resulting in bottlenecks and delivery delays~\cite{subbaswamy2022unifying}. 

Addressing data drifts is therefore essential for maintaining the high standards of quality necessary for effective ML applications. ML models must include robust monitoring and adaptation mechanisms to identify and rectify data drift promptly, ensuring the reliability and accuracy of their applications in a dynamic world.  To identify data drifts, both open-source and proprietary tools have been developed that implement various algorithms for detecting data drifts. Nevertheless, for ML engineers, selecting the most appropriate tool for a given ML application can be daunting. With an array of options, each with its unique strengths and limitations, making an informed choice requires careful consideration of the specific requirements and contexts of their projects. This complexity underscores the necessity for clear guidelines and comprehensive benchmarks to aid engineers in navigating the landscape of ML tools effectively.

Numerous surveys and benchmarks are available that assess and compare various detection algorithms \cite{bayram2022concept,gemaque2020overview,rabanser2019failing}. Nonetheless, a gap in the current literature is evident, as these benchmarks often fail to incorporate real-world datasets. They tend to focus narrowly on specific algorithms and fall short of evaluating tools designed for detection at an industrial scale. This limitation underscores the need for more comprehensive benchmarking that includes practical data and examines the full spectrum of detection tools suitable for large-scale applications.

To overcome this challenge, in this paper, we present a microbenchmark, denoted as \PaperAcronym{}\footnote{\PaperAcronym{} is an abbreviation of \textbf{D}ata \textbf{D}rift \textbf{D}etection \textbf{Bench}mark.}, that rigorously evaluates three leading open-source tools designed for the detection of data drifts, employing a range of established criteria. To conduct this assessment, we measure both functional and non-functional performance attributes. Our examination encompasses several key aspects: functional suitability, integration capability, compatibility with various data types, user-friendliness, and performance metrics including time efficiency and resource consumption. This holistic approach ensures a comprehensive analysis, providing valuable insights into the efficacy of each tool.

\PaperAcronym{} examines two distinct use cases, each using a univariate time series data set with building data from building management experts. These datasets comprise authentic data, where detecting data drifts poses a substantial challenge due to the data's intricate nature. The first use case delves into a dataset that encapsulates a concept shift within the dataset. Unlike the predictable nature of seasonal trends, concept shifts represent a change in the underlying relationships between data points, often resulting in a more elusive form of data drift that can be particularly challenging to detect and quantify. Transitioning to the second use case, we encounter a different kind of complexity: a data drift characterized by seasonal trends. The cyclical nature of this data offers a unique opportunity to explore how data drifts can manifest over time, influenced by predictable, recurring patterns. Since these datasets involve predictions for a single variable, \PaperAcronym{} focuses exclusively on univariate methods for detecting data drifts. Together, these two use cases provide a comprehensive view of the nuanced and multifaceted nature of data drifts, enabling a thorough comparison of the efficacy of our detection tools in real-world scenarios.

The remainder of the paper is structured as follows. Section~\ref{sec:preliminaries} delineates the various types of data drifts and introduces the architecture of \PaperAcronym{}. Section~\ref{sec:tools} describes the open-source detection tools that are the focus of our comparison. Subsequently, Section~\ref{sec:cases} details the two use cases deployed to examine the efficacy of these drift detection tools. Progressing to Section~\ref{sec:benchmark}, we engage in a comparative analysis of the tools across the use cases. 
Finally, Section~\ref{sec:conclusion} concludes the paper, offering perspectives on future work.
\section{Preliminaries \& Architecture}\label{sec:preliminaries}

\subsection{Types of Data Drifts}
Data drifts are categorized into four primary types: Covariate Drift, Prior Probability Drift, Concept Drift, and Dataset Shift~\cite{moreno2012unifying}. This classification hinges on two key considerations, including the nature and direction of change. First, it assesses whether the changes are occurring in the input variables, known as Covariate Drift, or in the target variable, referred to as Prior Probability Drift. Second, it explores the direction of influence between these variables. Specifically, it examines if the input variables are influencing the target variable ($X \rightarrow Y$ problem), or if the situation is reversed, with the target variable influencing the input variables ($Y \rightarrow X$ problem). The Concept Drift reflects a change in the \quotes{concept} so that the relationship between input variables and target variables changes. A Dataset Shift is present if there is neither a covariate shift, prior probability shift nor a concept shift and the joint distributions of target and input variables differ between the training and test sets.

Data drift can occur in various ways, impacting datasets either abruptly or subtly over time. An \textit{abrupt} data drift might arise when a new sensor with different calibration parameters is introduced, causing an immediate and noticeable change in measurements~\cite{gama2014survey}. This sudden variation can significantly affect data consistency and model performance. On the other hand, data drift can also emerge \textit{incrementally}, unfolding slowly as conditions change. For example, a sensor's accuracy might degrade over time, leading to a gradual deviation in the recorded measurements. This type of drift requires careful monitoring to detect and adjust for the changes, as they may not be immediately evident. Additionally, data drift can display periodic patterns, which we refer to as \textit{recurring} data drifts. These are often characterized by cyclical trends where certain patterns or concepts reemerge at regular intervals. Seasonal variations are a prime example, where the influence of time-related factors brings about predictable, recurring changes in the data. Understanding the nature of data drift—whether abrupt, incremental, or recurring—is crucial for maintaining the accuracy and reliability of ML models.

\subsection{Benchmark Overview}
In this section, the architecture of \PaperAcronym{} is delineated alongside its principal components. \PaperAcronym{} is designed to assess individual drift values, which in turn inform the functional suitability of the system. Additionally, it evaluates non-functional criteria such as runtime, CPU runtime, and memory consumption. At the core of \PaperAcronym{} lies the benchmark controller, the orchestrator that manages the various components of the benchmarking process. This controller is responsible for loading the datasets, which are derived from two distinct use cases, and for carrying out the necessary preprocessing steps. It systematically partitions the dataset into training and test sets, setting the stage for a thorough comparison of open-source drift detection tools based on pre-selected benchmarks. 
\begin{figure}[htbp]
    \centering
    \includegraphics[width=1\linewidth]{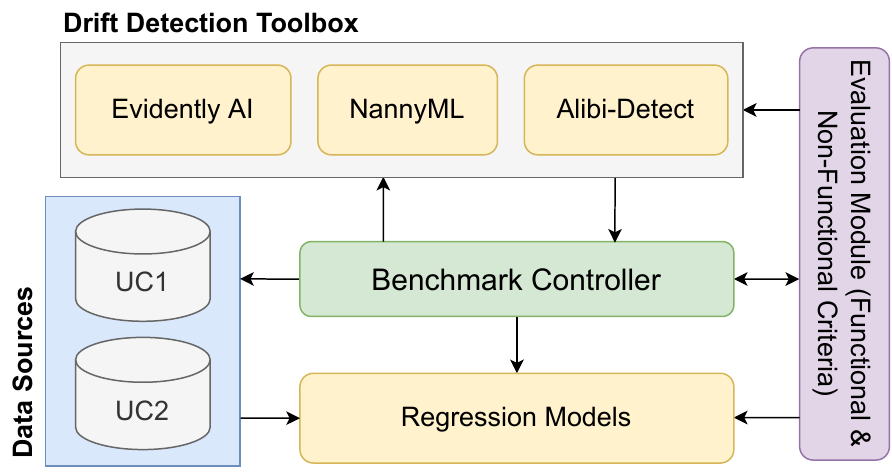}
    \caption{Architecture of \PaperAcronym{}}
    \label{fig:architecture}
\end{figure}

Within the scope of \PaperAcronym{}, we focus on three widely recognized open-source tools: Alibi-Detect, NannyML, and Evidently AI (cf. Section~\ref{sec:tools}). These tools, each with their unique methodologies, are applied independently to the dataset to ensure a comprehensive evaluation. The outcomes of the benchmarking process are documented. Results are not only displayed in the console for immediate review but are also stored in a CSV file, facilitating subsequent in-depth analysis. This methodical approach ensures that \PaperAcronym{} provides a robust framework for benchmarking drift detection tools, yielding insights that are both actionable and accessible.
%
\section{Drift Detection Tools}\label{sec:tools}
In this section, we introduce the three open-source drift detection tools evaluated in our study.
\paragraph{Evidently AI}
%
Evidently AI\footnote{https://docs.evidentlyai.com/} is a comprehensive open-source tool designed to rigorously test, monitor, and analyze ML models during their operational phase. It serves as a multifaceted suite within the realm of ML Operations (MLOps), encompassing functionalities such as Data Drift Detection, ML Monitoring, ML Quality Review, and Data Quality Assurance. For numerical variables, Evidently AI is equipped with an array of 12 distinct methods, each with its default threshold settings. One of the pivotal features of Evidently AI is its focus on univariate methods for the detection of data drifts, which necessitates the comparison of two datasets: a \textit{reference} dataset (usually the training data) and a \textit{current} dataset (akin to test data). For time series analysis, Evidently AI handles date-formatted variables automatically as time variables. Users also have the option to explicitly define time variables for enhanced clarity in visual representations, where these are plotted along the x-axis. However, the platform does not extend specialized processing options for time series data beyond this functionality.

Evidently AI promises to identify two principal forms of data drifts. including \textit{target variable} drift and \textit{input variable} drift. The former type of drift pertains to shifts in the target variable itself or in the predictions of the target variable. Even in the absence of actual measured values for the target variable, Evidently AI is capable of detecting data drifts through the predicted values. The second type is shifts in the input variables. Evidently AI's preconfigured methods are selected based on data characteristics, such as the size of the dataset, and the nature of the variables involved--be they numerical, categorical, or binary. Moreover, the choice of method may depend on the number of unique values a variable possesses. For example, Evidently AI recommends the Wasserstein Distance algorithm for numerical columns when the sample size exceeds 1000 observations.

Evidently AI's methods come with predefined threshold values, facilitating immediate utilization. Statistical tests within the tool are configured as one-sided, left-tailed, with a default lower threshold value of 0.05. An alarm is sounded if a test yields a drift value below this threshold, indicating significant deviation. For these tests, a p-value is typically provided, except in the case of Kullback-Leibler divergence, where the divergence value itself is reported. Conversely, the distance metrics are designed as one-sided, right-tailed tests with an established upper threshold of 0.1. Distances exceeding this value, such as a 0.2 result from the Wasserstein Distance measure, prompt an alarm, denoting a notable drift. With distance metrics, the tool reports the actual calculated distance. Evidently AI empowers users to calibrate alarm sensitivity via two distinct mechanisms: (1) the adjustment of method thresholds, and (2) modifying the proportion of variables that must exhibit drift before an alarm is triggered.

Evidently AI's report generation is a core feature, providing a succinct main view alongside a comprehensive detailed view. The main view offers calculated drift values for each input and the target variable to detect data drifts. Should shifts be observed in over 50\% of the variables, an alarm is triggered, signaling broad data changes. Users can tailor the drift detection threshold, customizing when the alarm activates. High-impact variables can be given greater weight, making the system particularly responsive to their shifts. The detailed view augments this with visual representations of drift and comparative distribution charts for training and test datasets, enabling an in-depth analysis of individual variables.

\paragraph{NannyML}
NannyML, an open-source Python library\footnote{https://nannyml.readthedocs.io/en/stable/}, excels in evaluating model performance, identifying data drifts, and estimating business value, all while ensuring data quality. It accommodates univariate and multivariate detection methods and facilitates insightful visualizations for tabular data. The library effectively handles classification and regression problems. Additionally, it seamlessly integrates visual indicators of data drifts with corresponding model performance metrics, enabling a clear understanding of their impact. NannyML detects data drifts by juxtaposing two distinct datasets: the baseline (training data) and the current dataset (test data). It segments the latter into \quotes{chunks}, which could be individual data points or aggregated batches. For time series analysis, it allows for \quotes{time-based chunking}, slicing data by specific periods. Alternatively, \quotes{size-based chunking} splits by the count of observations, while \quotes{number-based chunking} divides into a predefined quantity of chunks, with ten being the default setting.

NannyML delineates the timeframe of each data chunk, providing start and end dates, and flags any detected data drifts, enabling precise, timely drift alarms. The platform incorporates four detection algorithms for continuous univariate data, each backed by configurable threshold values that bracket expected behavior. Should computed values breach these boundaries—exceeding the upper or falling below the lower limit—a drift alarm is activated. NannyML features two threshold paradigms: Constant Thresholds, which are not preset but can be specified with upper and lower bounds, and Standard Deviation Thresholds, which pivot around the training data's mean and scale with a factor—defaulting to three—applied to the standard deviation, forming the basis for the drift criteria.

NannyML informs users about appropriate method applications and their unique attributes. For 
continuous data, it advises employing either the Jensen-Shannon Distance or Wasserstein Distance. The Jensen-Shannon Distance is well-suited for spotting minor data drifts, albeit with an increased likelihood of false positives. Conversely, the Wasserstein Distance is particularly sensitive to outliers, producing a metric that spans from zero to infinity, reflecting the degree of detected shifts. NannyML extends its utility by enabling users to analyze and visualize a model's predictive accuracy. It uniquely offers accuracy estimations for ML models even in the absence of ground truth data. To this end, NannyML introduces two methodologies: Confidence-based Performance Estimation (CBPE) leverages the confidence scores of predictions for application in classification models, while Direct Loss Estimation (DLE) is designed for regression models, quantifying predictive error directly. These tools enhance the ability to gauge model reliability under conditions where traditional evaluation metrics cannot be calculated.

\paragraph{Alibi-Detect}
Alibi-Detect\footnote{https://docs.seldon.io/projects/alibi-detect/en/stable/}, an open-source Python library, specializes in identifying outliers, adversarial attacks, and data drifts. It supports multiple backends such as PyTorch, TensorFlow, KeOps, and Prophet, and can function independently of these platforms. Method availability is backend-dependent, with TensorFlow and Prophet providing the broadest selection. Drift detection in Alibi-Detect is segmented into online and offline techniques, along with univariate and multivariate strategies. Some techniques necessitate the use of TensorFlow or PyTorch for classification or regression models. In \PaperAcronym{}, we focus exclusively on offline methods that do not rely on a predetermined model. Alibi-Detect's univariate options for numerical data are cataloged, operating solely on statistical tests with a standard significance threshold of 0.05.

Alibi-Detect outputs a binary value for each drift detection task, signifying whether a divergence (True) or no divergence (False) exists between the compared distributions. This binary indicator serves as a trigger for an alarm in the event of a detected data drift in the scrutinized variable. Capable of identifying shifts in both complete datasets and individual variables, Alibi-Detect offers precision in localizing changes. The Spot-The-Difference Test, an Alibi-Detect innovation inspired by Jitkrittum et al. [20], exemplifies the library's capability. This classifier-based test trains a model on a data subset to differentiate between training and test datasets; successful differentiation suggests the occurrence of a shift.

\paragraph{Data Drift Methods}

The data drift detection capabilities of the compared tools are summarized in Table \ref{tab:comparison-methods}, which compares the methods each tool provides. It is crucial to highlight that although some methods are common across multiple tools, their implementations can differ significantly. Each tool may employ its unique algorithmic interpretation or configuration, which can result in varying drift metrics. This diversity in implementation underlines the need for careful consideration when interpreting and comparing the drift values produced by different tools. For instance, while the Kolmogorov-Smirnov Test may be available across Evidently AI, NannyML, and Alibi-Detect, the specificities of each tool's approach—such as the handling of data preprocessing, the computation of test statistics, or the calibration of significance levels—can lead to distinct outcomes. Consequently, practitioners should not assume uniformity in results across different platforms, even when the underlying method bears the same name.
\begin{table}[h]
\centering
\caption{Data drift detection methods in each tool}
\label{tab:comparison-methods}
\resizebox{\columnwidth}{!}{%
\begin{tabular}{lccc}
\toprule
\textbf{Method}             & \textbf{Evidently AI} & \textbf{NannyML} & \textbf{Alibi-Detect} \\ \midrule
Anderson Darling Test       & \checkmark            & —                & —                     \\ 
Cramér-von-Mises Test       & \checkmark            & —                & \checkmark            \\ 
Energy Distance             & \checkmark            & —                & —                     \\ 
Epps-Singleton Test         & \checkmark            & —                & —                     \\ 
Hellinger Distance          & \checkmark            & \checkmark       & —                     \\ 
Jensen-Shannon Distance     & \checkmark            & \checkmark       & —                     \\ 
Kolmogorov-Smirnov Test     & \checkmark            & \checkmark       & \checkmark            \\ 
Kullback-Leibler Divergence & \checkmark            & —                & —                     \\ 
Mann-Whitney U-Rank Test    & \checkmark            & —                & —                     \\ 
Population Stability Index  & \checkmark            & —                & —                     \\ 
Spot-The-Difference         & —                     & —                & \checkmark            \\ 
T-Test                      & \checkmark            & —                & —                     \\ 
Wasserstein Distance        & \checkmark            & \checkmark       & —                     \\ \bottomrule
\end{tabular}%
}
\end{table}

%
\section{Real-world Use Cases}\label{sec:cases}
In this section, we present two real-world use cases within the field of building management. Each case involves the analysis of time series data that exhibits data drift over time.
%
\subsection{UC1: Occupancy Detection}
%
The first use case examines the temporal pattern of room occupancy, distinguishing between states of being occupied by one or more individuals and being unoccupied. The dataset encompasses 46,555 time series entries, spanning from March 30, 2021, at 12:12 AM to July 11, 2021, at 11:57 PM, with data recorded at 3-minute intervals. The dataset includes air quality parameters of a room as variables. It measures the room temperature (\texttt{temperature}) and the current CO2 level of the room (\texttt{co2}). A Deep Learning (DL) classification model predicts based on these variables whether a room is occupied (value: 1) or not (value: 0). At the same time, the ground truth is available that describes the actual reality of whether the room is occupied or not. The classification model has been trained on data from a room in another building and improved by transfer learning with a sample from the room under investigation from March 22, 2021, to March 26, 2021.

\begin{figure}[htbp]
    \centering
    \includegraphics[width=\linewidth]{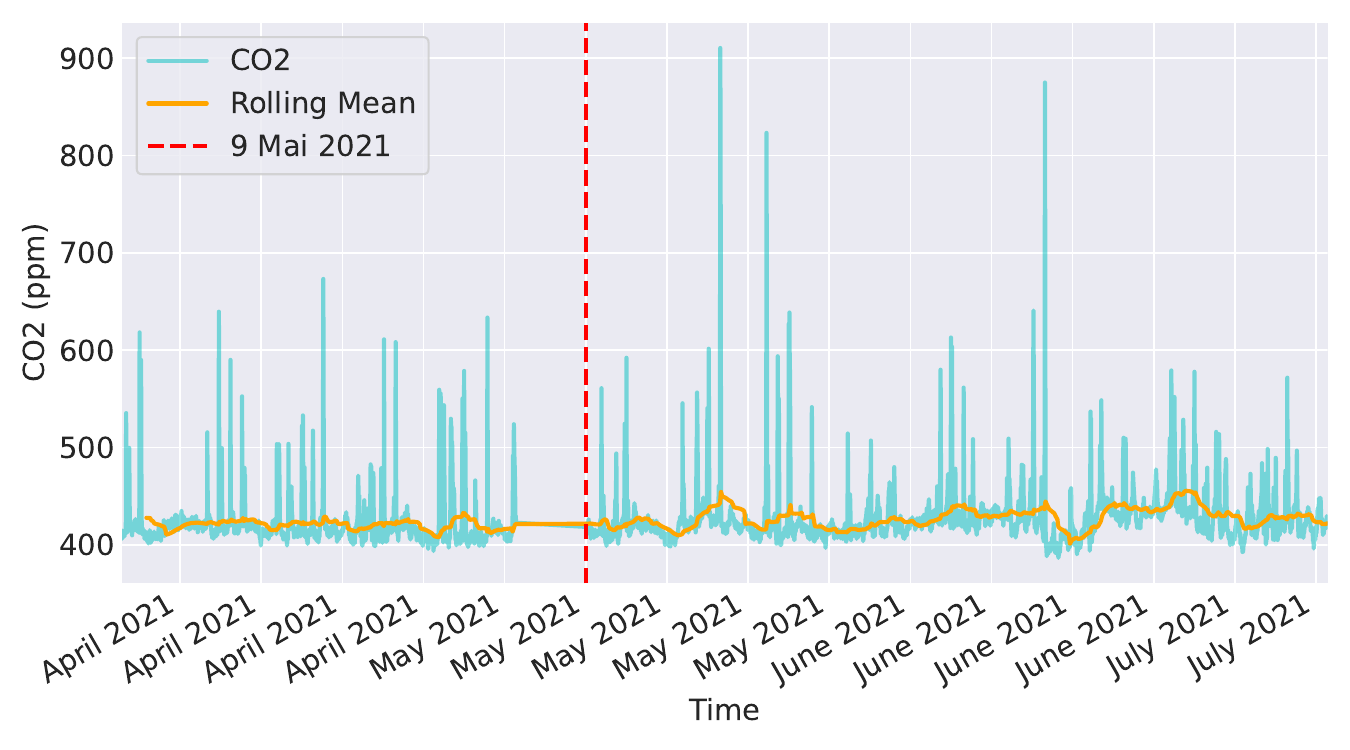}
    \caption{Development of CO2 over time}
    \label{fig:co2}
\end{figure}

The data drift analysis encompasses the input variables \texttt{co2} and room \texttt{temperature}, alongside the target variable  \texttt{measured} throughout the period between March 30 and July 11, 2021. Notably, there is a data gap from April 30 to May 9, 2021 (cf. Figure~\ref{fig:co2}). It is after this gap, starting May 9, 2021, that the ventilation system's operation shifted from constant to variable based on real-time air quality metrics. This shift induces recognizable patterns of fluctuation in the dataset, differentiating pre-May 9 data (training dataset) from post-May 9 data (test dataset). Figure~\ref{fig:co2} shows that three outliers with CO2 concentrations exceeding 800 ppm are observed, along with several instances below 400 ppm during the week of June 20. These larger fluctuations suggest a data drift in the CO2 variable between the training and test datasets. Additionally, the presence of a shift can be inferred from the test data's higher or lower values relative to the training data. Consequently, the dataset is divided into two parts: training data, which includes records from March 30, 2021, at 12:12 AM to May 8, 2021, at 11:57 PM, and test data, which spans from May 9, 2021, at 12:00 AM to July 11, 2021, at 11:57 PM.

\begin{figure}[htbp]
    \centering
    \includegraphics[width=\linewidth]{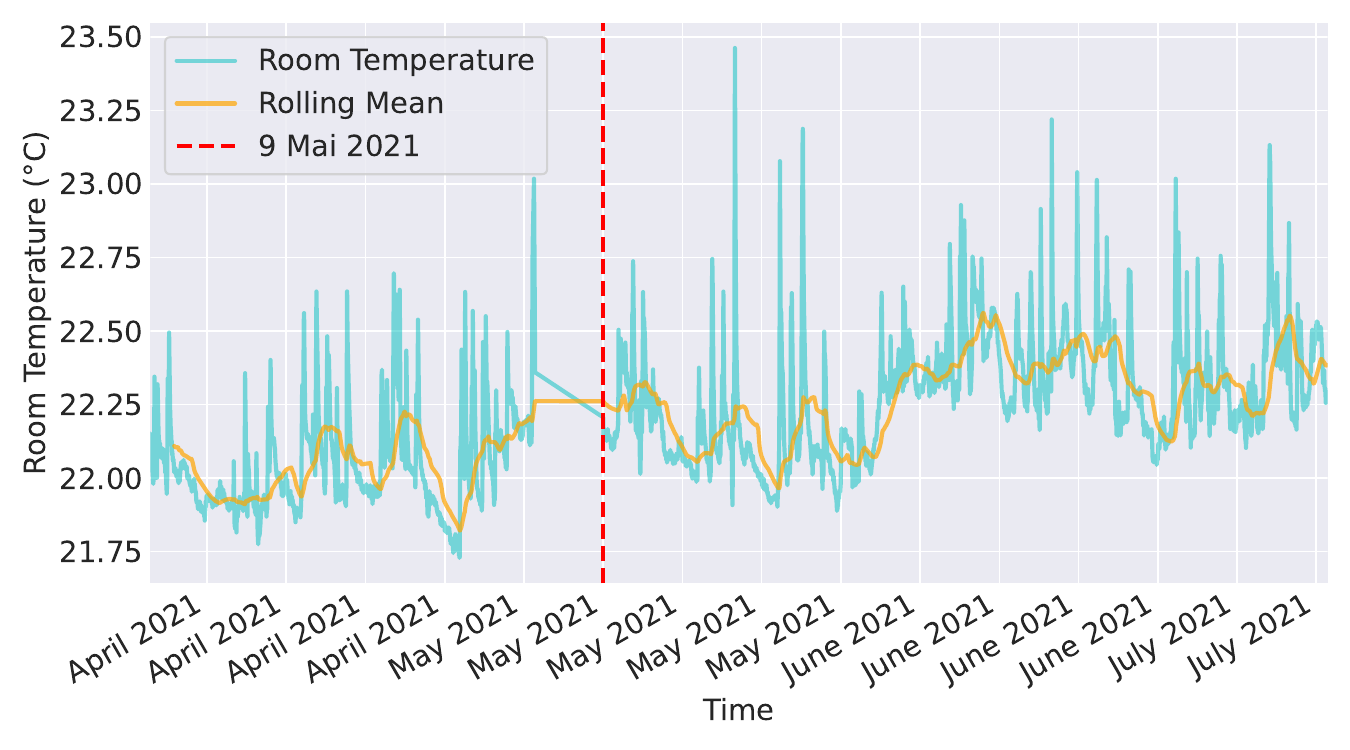}
    \caption{Development of room temperature over time}
    \label{fig:room_temp}
\end{figure}

\begin{figure}[htbp]
    \centering
    \includegraphics[width=\linewidth]{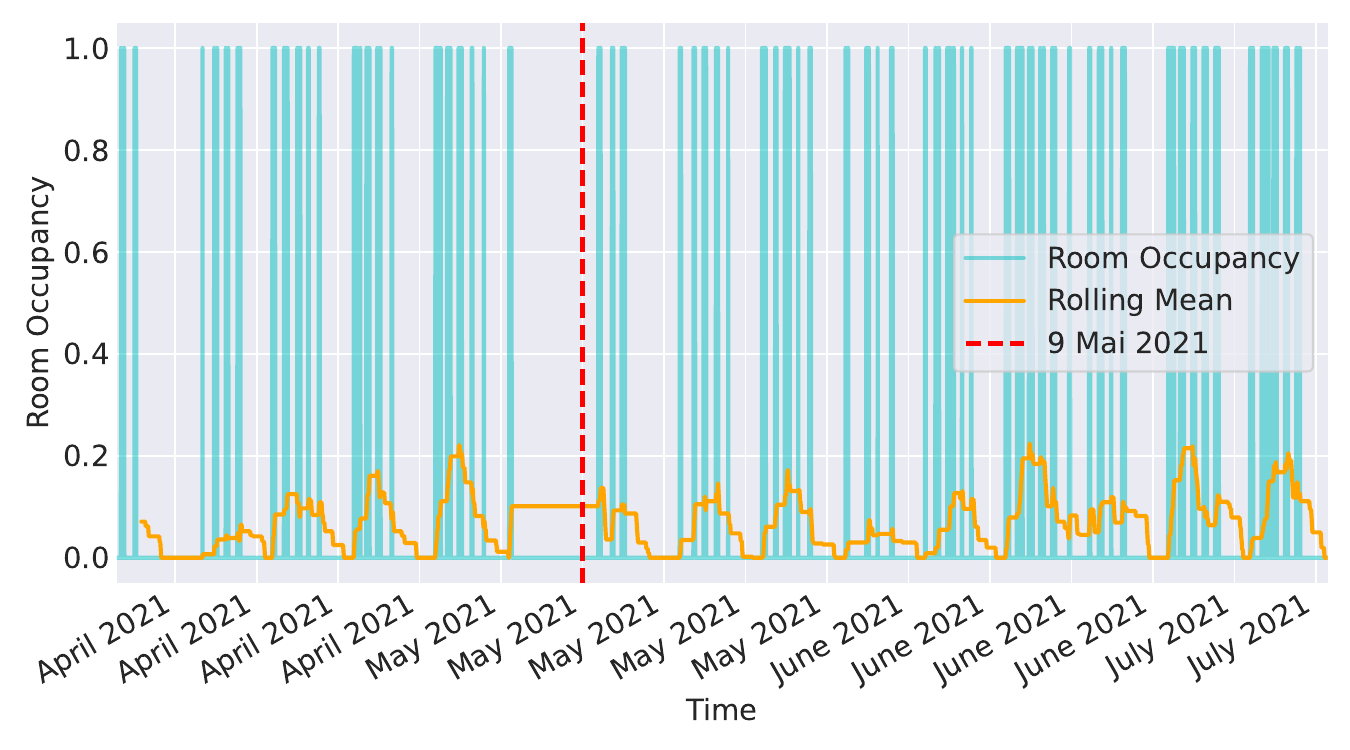}
    \caption{Development of occupancy over time}
    \label{fig:occupancy}
\end{figure}

Figure~\ref{fig:room_temp} depicts the room temperature's progression in \( ^\circ\text{C} \) throughout the test period, revealing a noticeable increase compared to the training data and indicating a drift in the temperature variable. Figure~\ref{fig:occupancy} illustrates the temporal pattern of room occupancy during the test period. The target variable's distribution remains stable, exhibiting no significant fluctuations compared to the training data, suggesting an absence of data drift. In summary, there is a recognizable data drift in the input variables $x$, but no such drift is detected in the target variable $y$. Given that $P_{tr}(y) = P_{tst}(y)$ and $P_{tr}(x|y) \neq P_{tst}(x|y)$, the dataset for this use case exhibits a concept shift.


\subsection{UC2: Energy Consumption Prediction}
%
The second use case describes the temporal evolution of thermal energy usage across a network of 36 university buildings, offering a comprehensive dataset encompassing 1,556,915 individual time series entries. These entries log thermal energy usage, quantified in megawatt-hours (MWh), and span from the early hours of January 1, 2019, at 01:00 AM, to the end of the recorded period on June 1, 2022, at 12:00 AM. Situated in Tampere, Finland, these buildings are integral components of the local district heating system. To monitor thermal energy usage, each building is equipped with a primary heat meter. These meters are either advanced smart meters, which offer real-time data and remote monitoring capabilities, or traditional analog metering devices, which require manual reading. Regardless of the type, these meters log the daily energy consumption and estimate the hourly usage by computing the difference between the cumulative readings of consecutive hours.

\sloppy The time series dataset comprises the following variables: \texttt{consumption} (measured thermal energy consumption in MWh), \texttt{ids} (building ID), \texttt{time} (timestamp), and \texttt{temp\_outside} (i.e., outside temperature of the nearest weather station in \( ^\circ\text{C} \)). For data drift analysis, we focus on data pertaining to a building identified as Building ID 1, spanning from April 1, 2019, at 12:00 AM, to March 31, 2022, at 11:00 PM. During this analysis, two variables are examined: the target variable \texttt{consumption} and the input variable \texttt{temp\_outside}. To align with the timeline of this study, the training data is chosen to encompass an entire year, initiating at the start of April 1, 2019, at 12:00 AM, and concluding on March 31, 2020, at 11:00 PM. Meanwhile, the test data is compiled to include the subsequent two-year timeframe, extending from April 1, 2020, at 12:00 AM, to March 31, 2022, at 11:00 PM. This temporal division ensures a comprehensive and robust examination of the potential drifts between the training and testing stages of the dataset.

\begin{figure}[htbp]
    \centering
    \includegraphics[width=\linewidth]{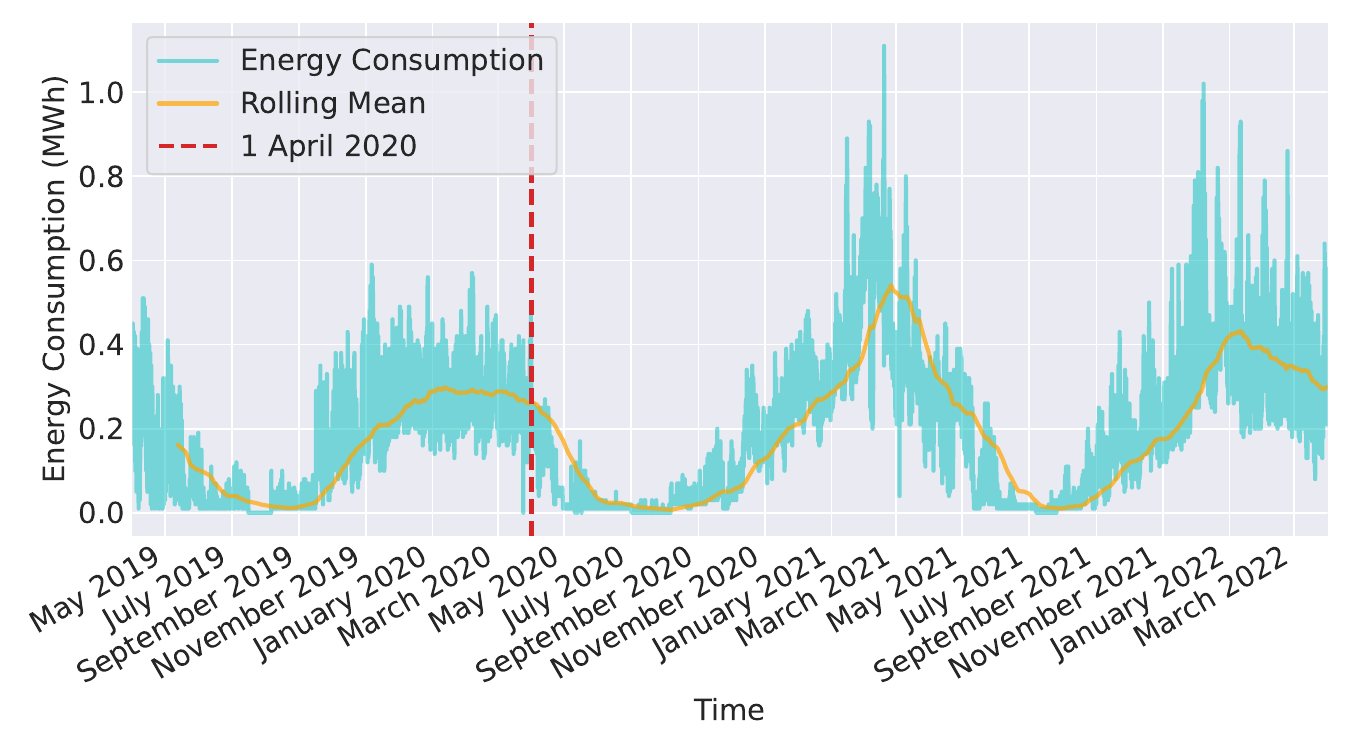}
    \caption{Development of energy consumption over time}
    \label{fig:energy}
\end{figure}

Through comparing test to training segments, an evidence of data drift across their attributes is revealed. Specifically, a notable drift is observed in the energy consumption metrics. Figure~\ref{fig:energy} illustrates the temporal pattern of energy consumption throughout the test period, with a demarcating red line distinguishing the training data on the left from the test data on the right. During the testing phase, the average consumption escalates from 106.5 kWh to 111.7 kWh, a transition that is also characterized by a more pronounced occurrence of outliers compared to the training data (especially during the winter months). Moreover, the test set reflects an obvious drift in the average outside temperature, recording a cooler mean of 5.55\( ^\circ\text{C} \) against the training set's warmer average of 6.24\( ^\circ\text{C} \). This divergence in the distribution of both the target variable \( y \) and the input variable \( x \) between the training and test sets is indicative of Dataset Shift, as described by the following probabilistic disparities: \( P_{tr}(x) \neq P_{tst}(x) \), \( P_{tr}(y|x) \neq P_{tst}(y|x) \), and \( P_{tr}(y,x) \neq P_{tst}(y,x) \). 
%
%


%
\section{Comparative Analysis}\label{sec:benchmark}
%
In this section, we first outline the framework of our evaluations, and then we proceed to detail the findings from our comparative study.

\subsection{Experimental Setup}
%
\textbf{Preprocessing:} Data preprocessing is a critical initial step for both use cases, beginning with universal procedures such as loading CSV files, purging duplicates to ensure data uniqueness, converting time columns into \texttt{datetime} format for proper temporal analysis, and omitting missing values. After these foundational steps, specific preprocessing techniques are then applied to each use case as required. For Use Case 1 (UC1), we executed a polynomial interpolation on the CO2 and temperature variables to estimate unknown or missing values within the time series through polynomial functions. This was succeeded by the application of a Kalman smoother, technique employed to refine past noisy measurements by incorporating current observed values. These preprocessing steps were implemented to condition the data for subsequent predictions with the DL classification model.

For Use Case 2 (UC2), each building has been initially anonymized with a unique identifier. Data series lacking an ID were deemed irrelevant and thus excluded from the analysis. Furthermore, we examined high energy consumption readings; instances where energy values were absent for several hours and subsequently spiked were disregarded, as these peaks result from cumulative calculations since the last recorded value. Additionally, negative energy consumption figures were omitted from our analysis, as they generally suggest measurement inaccuracies. While negative readings could theoretically indicate a surplus of generated energy over consumption, such scenarios are implausible during the Finnish winter and are not considered valid in this context.

\textbf{Moving Average \& Growth Rate:} To justify the adoption of data drift detection tools, a moving average approach from time series analysis is first utilized. For UC1, where no clear periodicity exists, a smaller window of \( p = 3 \) is selected. For the training data in UC2, a window of \( p = 2196 \) values is chosen to represent a quarter, accounting for a leap year, while the test data uses \( p = 2190 \). This smoothes the data, capturing quarterly trends. The resulting smoothed series maintains recognizable patterns, as demonstrated in Figures~\ref{fig:co2}-\ref{fig:energy}. Using the computed moving averages, the growth rates for the training and test data are determined. In UC1, no significant growth or decline in the variables is revealed, with the target variable's growth rate at 1.0938, and the input variables displaying more modest rates of 1.0107 for temperature and 1.0157 for the CO2 variable. In contrast, the growth rates for both the target and input variables in UC2 indicate a shift in the data: the target variable exhibits an increase by a factor of 1.2921, while the input variable's growth rate is 1.2104. 

\textbf{Evaluation Metrics}: \PaperAcronym evaluates the compared open-source tools using four functional and two non-functional metrics. The first functional criterion is the ability of a tool to identify data drifts accurately, assessed using the number of methods it incorporates for accurate detections. This is crucial for the deployment of such tools in a production environment, where the reliability of detecting genuine data drifts is paramount.  A visual representation can significantly aid in the interpretation of these drifts. Moreover, the tool should trigger an alarm upon detecting a drift, providing a precise timestamp. This enables a comparison with the model's forecast accuracy.

To clearly distinguish between the capabilities of various methods and tools for data drift detection, we classify the methods into four groups based on their detection accuracy: Group One includes methods that correctly identify shifts in both input and target variables; Group Two comprises methods that recognize shifts in input variables only; Group Three consists of methods that detect shifts solely in the target variable; and Group Four encompasses methods that do not correctly detect shifts in either. The second functional criterion evaluates the tool's potential for integration into a production environment. Effective integration options are vital for incorporating a detection tool into existing MLOps workflows. The third functional criterion, flexibility, assesses the variety of data types the tool can handle. For sustainability and adaptability in long-term usage, it is imperative for the tool to support a range of data types beyond numerical values, including text and image data. The final functional criterion focuses on user-friendliness. This involves examining the experiences encountered while connecting to and utilizing the tools.

Non-functional criteria address the quantitative attributes demanded of a system, focusing on performance aspects like temporal efficiency and resource usage. The first non-functional criterion evaluates the tool's runtime. When utilizing detection tools in an industrial context, it is paramount that a data drift is identified with minimal delay to ensure timely responses and maintain operational continuity. The time behavior of a tool is quantified using two metrics: Variant 1 measures the total runtime including any waiting time for resources, and Variant 2 measures the CPU runtime excluding waiting times. Both variants start by recording the current time before and after the execution of the detection function. The runtime is calculated as the difference between the end and start times, divided by the tool's number of methods, yielding the average runtime per method in milliseconds. Finally, the second non-functional criterion assesses memory usage, emphasizing the importance of not only rapid data drift detection but also minimal use of resources. We run all the experiments on an Ubuntu 20.04 LTS machine with 6 2.60 GHz cores and 30 GB memory. 

\subsection{Results of UC1}
%
Table~\ref{tab:uc1_results} summarizes the estimated drift values from the three open-source tools in UC1. Below, we present the results of each tool.

\begin{figure}[htbp]
	\centering
	\subfloat[\texttt{CO2} variable]{\label{fig:evidently_co2}\includegraphics[width=\columnwidth]{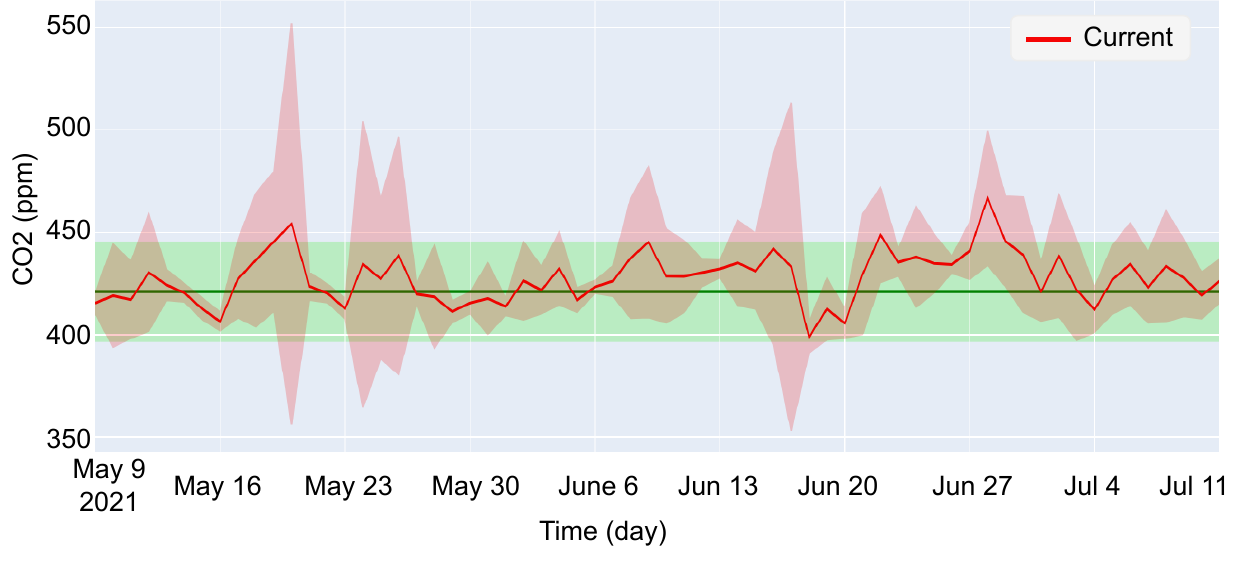}} \newline
	\subfloat[\texttt{Occupancy} variable]{\label{fig:occupancy_uc1}\includegraphics[width=\columnwidth]{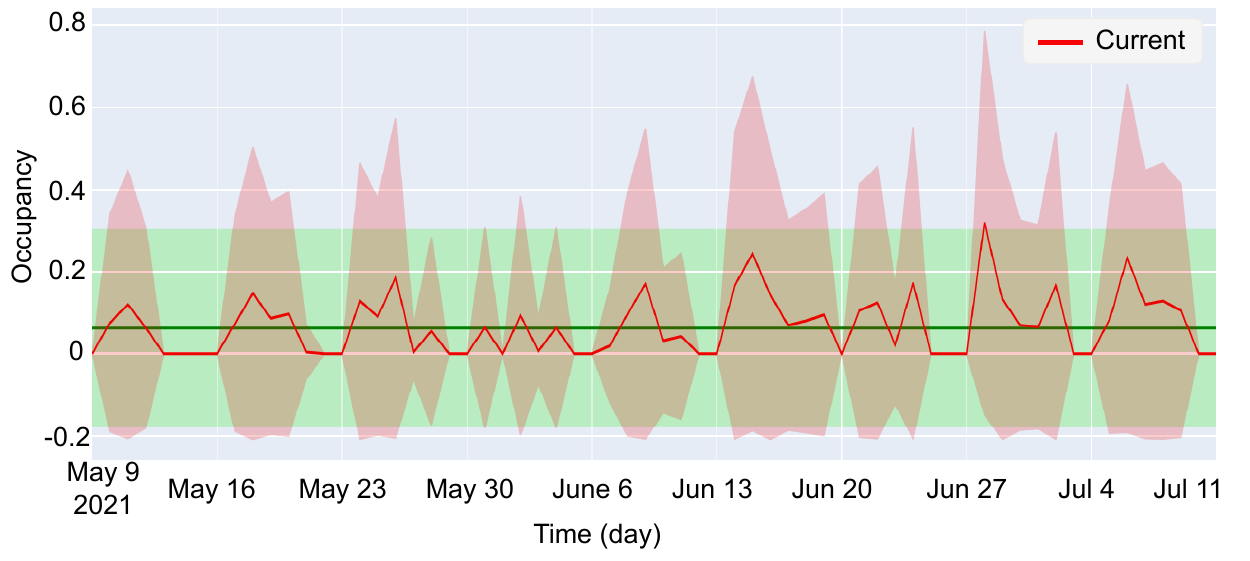}} \newline
    \subfloat[Distribution of \texttt{Temperature} variable]{\label{fig:temperature_uc1}\includegraphics[width=\columnwidth]{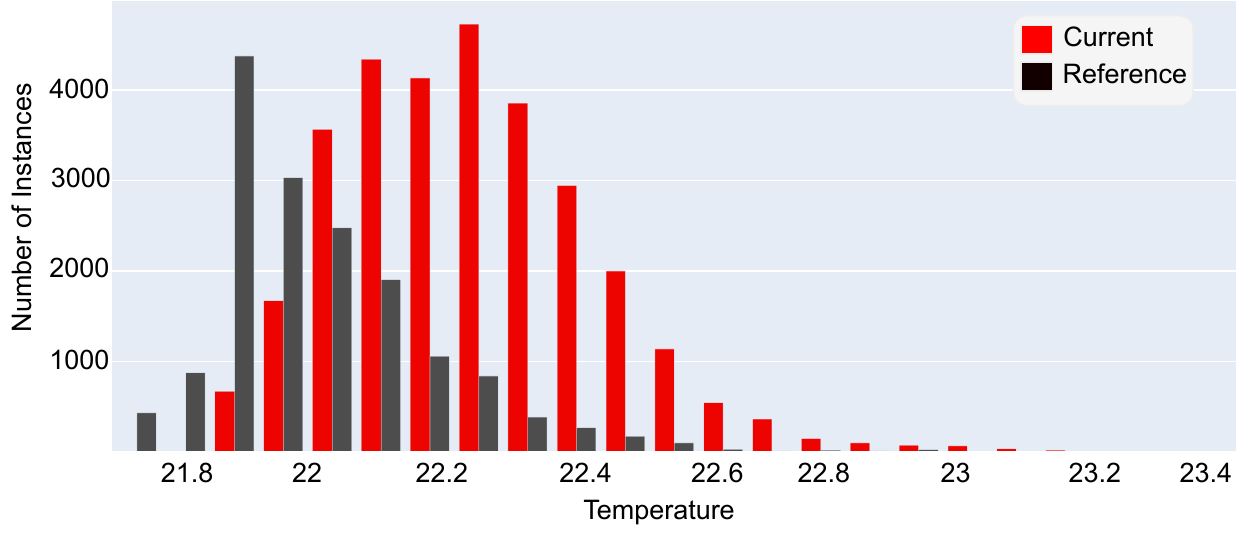}}
	\caption{Reports generated by Evidently AI for UC1}
	\label{fig:evidently_results_uc1} 
\end{figure}

\begin{table*}[htbp]
\centering
\caption{UC1 results using three open-source tools, indicating correct decisions with \ding{55} for Occupancy and \checkmark for other variables.}
\label{tab:uc1_results}
\resizebox{\textwidth}{!}{%
\begin{NiceTabular}{@{}lccccccccccccccccccccc@{}}
\toprule
\textbf{Tool} &
  \multicolumn{7}{c}{\textbf{Evidently AI}} &
  \multicolumn{7}{c}{\textbf{NannyML}} &
  \multicolumn{7}{c}{\textbf{Alibi-Detect}} \\ \midrule
Variable &
  \multicolumn{2}{c}{Occupancy} &
  \multicolumn{2}{c}{CO2} &
  \multicolumn{2}{c}{Temperature} &
  \multicolumn{1}{l}{} &
  \multicolumn{2}{c}{Occupancy} &
  \multicolumn{2}{c}{CO2} &
  \multicolumn{2}{c}{Temperature} &
  \multicolumn{1}{l}{} &
  \multicolumn{2}{c}{Occupancy} &
  \multicolumn{2}{c}{CO2} &
  \multicolumn{2}{c}{Temperature} &
  \multicolumn{1}{l}{} \\ \cmidrule(r){1-7} \cmidrule(lr){9-14} \cmidrule(lr){16-21}
Methods &
  \multicolumn{1}{l}{D-value} &
  \multicolumn{1}{l}{Drift?} &
  \multicolumn{1}{l}{D-value} &
  \multicolumn{1}{l}{Drift?} &
  \multicolumn{1}{l}{D-value} &
  \multicolumn{1}{l}{Drift?} &
  \multicolumn{1}{l}{Group} &
  D-value &
  Drift? &
  D-value &
  Drift? &
  D-value &
  Drift? &
  \multicolumn{1}{l}{Group} &
  D-value &
  Drift? &
  D-value &
  Drift? &
  D-value &
  Drift? &
  Group \\ \midrule
Kolmogorov-Smirnov Test &
  0,86 &
  \ding{55} &
  0,0 &
  \checkmark &
  0,0 &
  \checkmark &
  1 &
  0,01 &
  \ding{55} (0\%) &
  0,30 &
  \ding{55} (20\%) &
  0,62 &
  \ding{55} (20\%) &
  3 &
  0,86 &
  \ding{55} &
  0,0 &
  \checkmark &
  0,0 &
  \checkmark &
  1 \\
Wasserstein Distance &
  0,02 &
  \ding{55} &
  0,27 &
  \checkmark &
  1,35 &
  \checkmark &
  1 &
  0,01 &
  \ding{55} (0\%) &
  9,31 &
  \ding{55} (10\%) &
  0,23 &
  \checkmark (50\%) &
  3 &
  \Block[fill=gray!15]{1-7}{} &
   &
   &
   &
   &
   &
   \\
Kullback-Leibler Divergence &
  $2,76e^{-4}$ &
  \ding{55} &
  0,09 &
  \ding{55} &
  1,11 &
  \checkmark &
  3 &
  \Block[fill=gray!15]{1-14}{} &
   &
   &
   &
   &
   &
   &
   &
   &
   &
   &
   &
   &
   \\
Population Stability Index &
  $5,6e^{-4}$ &
  \ding{55} &
  0,18 &
  \checkmark &
  1,90 &
  \checkmark &
  1 &
  \Block[fill=gray!15]{1-14}{} &
   &
   &
   &
   &
   &
   &
   &
   &
   &
   &
   &
   &
   \\
Jensen-Shannon Distance &
  0,459 &
  \ding{55} &
  0,15 &
  \checkmark &
  0,43 &
  \checkmark &
  1 &
  0,03 &
  \ding{55} (0\%) &
  0,33 &
  \checkmark (100\%) &
  0,63 &
  \checkmark (100\%) &
  1 &
  \Block[fill=gray!15]{1-7}{} &
   &
   &
   &
   &
   &
   \\
Anderson Darling Test &
  $1,19e^{-3}$ &
  \checkmark &
  $1,00e^{-3}$ &
  \checkmark &
  $1,00e^{-3}$ &
  \checkmark &
  2 &
  \Block[fill=gray!15]{1-14}{} &
   &
   &
   &
   &
   &
   &
   &
   &
   &
   &
   &
   &
   \\
Cramér-von-Mises Test &
  $2,53e^{-7}$ &
  \checkmark &
  $7,08e^{-8}$ &
  \checkmark &
  $2,87e^{-7}$ &
  \checkmark &
  2 &
  \Block[fill=gray!15]{1-7}{} &
   &
   &
   &
   &
   &
   &
  $2,53e^{-7}$ &
  \checkmark &
  $2,87e^{-7}$ &
  \checkmark &
  $7,08e^{-8}$ &
  \checkmark &
  2 \\
Hellinger Distance &
  $8,36e^{-3}$ &
  \ding{55} &
  0,15 &
  \checkmark &
  0,45 &
  \checkmark &
  1 &
  0,02 &
  \ding{55} (0\%) &
  0,29 &
  \checkmark (100\%) &
  0,60 &
  \checkmark (100\%) &
  1 &
  \multicolumn{1}{l}{\Block[fill=gray!15]{1-7}{}} &
  \multicolumn{1}{l}{} &
  \multicolumn{1}{l}{} &
  \multicolumn{1}{l}{} &
   &
   &
   \\
Mann-Whitney U-Rank Test &
  $7,92e^{-3}$ &
  \checkmark &
  0,0 &
  \checkmark &
  0,0 &
  \checkmark &
  2 &
  \Block[fill=gray!15]{1-14}{} &
   &
   &
   &
   &
   &
  \multicolumn{1}{l}{} &
  \multicolumn{1}{l}{} &
  \multicolumn{1}{l}{} &
  \multicolumn{1}{l}{} &
  \multicolumn{1}{l}{} &
   &
   &
   \\
Energy Distance &
  $8,25e^{-3}$ &
  \ding{55} &
  1,45 &
  \checkmark &
  0,41 &
  \checkmark &
  1 &
  \Block[fill=gray!15]{1-14}{} &
   &
   &
   &
   &
   &
  \multicolumn{1}{l}{} &
  \multicolumn{1}{l}{} &
  \multicolumn{1}{l}{} &
  \multicolumn{1}{l}{} &
  \multicolumn{1}{l}{} &
   &
   &
   \\
Epps-Singleton Test &
  -- &
  -- &
  -- &
  -- &
  -- &
  -- &
  -- &
  \Block[fill=gray!15]{1-14}{} &
  \multicolumn{1}{l}{} &
  \multicolumn{1}{l}{} &
  \multicolumn{1}{l}{} &
  \multicolumn{1}{l}{} &
  \multicolumn{1}{l}{} &
  \multicolumn{1}{l}{} &
  \multicolumn{1}{l}{} &
  \multicolumn{1}{l}{} &
  \multicolumn{1}{l}{} &
  \multicolumn{1}{l}{} &
   &
   &
   \\
T-Test &
  0,01 &
  \checkmark &
  \multicolumn{1}{l}{$5,99e^{-129}$} &
  \checkmark &
  0,0 &
  \checkmark &
  2 &
  \Block[fill=gray!15]{1-14}{} &
   &
   &
   &
   &
   &
   &
  \multicolumn{1}{l}{} &
  \multicolumn{1}{l}{} &
  \multicolumn{1}{l}{} &
  \multicolumn{1}{l}{} &
   &
   &
   \\
Spot-The-Difference Test &
  \Block[fill=gray!15]{1-14}{} &
  \multicolumn{1}{l}{} &
  \multicolumn{1}{l}{} &
  \multicolumn{1}{l}{} &
  \multicolumn{1}{l}{} &
  \multicolumn{1}{l}{} &
  \multicolumn{1}{l}{} &
  \multicolumn{1}{l}{} &
  \multicolumn{1}{l}{} &
  \multicolumn{1}{l}{} &
  \multicolumn{1}{l}{} &
  \multicolumn{1}{l}{} &
  \multicolumn{1}{l}{} &
  \multicolumn{1}{l}{} &
  \multicolumn{1}{l}{$3,63e^{-118}$} &
  NA &
  \multicolumn{1}{l}{$3,63e^{-118}$} &
  NA &
  $3,63e^{-118}$ &
  NA &
  NA \\ \cmidrule(r){1-22}
\end{NiceTabular}%
}
\end{table*}
\textbf{Evidently AI: } For Evidently AI, the Epps-Singleton Test does not yield results for the variables considered, so the remaining eleven methods are examined. Among these, the majority of the methods (7 out of 11) do not detect a data drift in the \texttt{occupancy} variable. However, a data drift in the variables \texttt{CO2} (detected by 10 out of 11 methods) and \texttt{temperature} (detected by all 11 methods) is reliably identified. Evidently AI's report provides a detailed analysis for both input and target variables. Figure~\ref{fig:evidently_co2} highlights the data drift in the CO2 variable using Wasserstein Distance, with the measured value exceeding the standard deviation at three specific time points. 

For the target variable, Figure~\ref{fig:occupancy_uc1} shows that the occupancy readings consistently cluster near the mean and within one standard deviation. Only once, in the week from June 27th to July 4th, does a value momentarily breach the standard deviation boundary. However, this outlier does not cross the 0.1 threshold, leading Evidently AI to conclude, via Wasserstein Distance, that there is no detectable drift in the data. Figure~\ref{fig:temperature_uc1} compares the temperature variable's training data (in red) with the test data (in black), revealing a noticeable divergence. The test data exhibits higher temperatures, shifting the distribution visibly to the right. This directional drift is readily apparent in the frequency plot, highlighting a significant deviation between the datasets. 

\textbf{NannyML:} NannyML identifies data drift in the \texttt{temperature} variable using three of four methods for over 50\% of the chunks (cf. Table~\ref{tab:uc1_results}). For the \texttt{CO2} variable, two methods detect drift in more than 50\% of the chunks. Meanwhile, no drift is detected in the target variable. In NannyML, both the Jensen-Shannon and Hellinger Distances accurately detect data drift in the input and target variables. Conversely, the Kolmogorov-Smirnov Test and Wasserstein Distance fail to identify shifts in the input variables for most data segments, yielding inaccurate assessments. However, they accurately confirm the absence of drift in the target variable.
\begin{figure}[htbp]
	\centering
	\subfloat[\texttt{CO2} variable]{\label{fig:nannyml_co2_uc1}\includegraphics[width=\columnwidth]{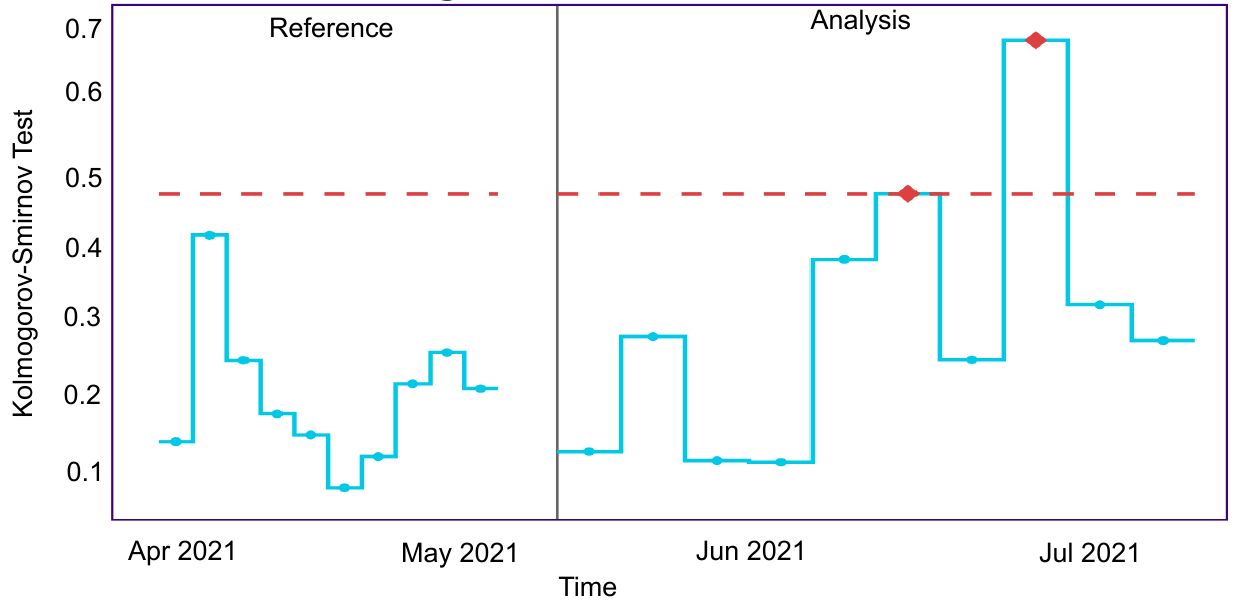}} \newline
	\subfloat[\texttt{Temperature} variable]{\label{fig:nannyml_temp_uc1}\includegraphics[width=\columnwidth]{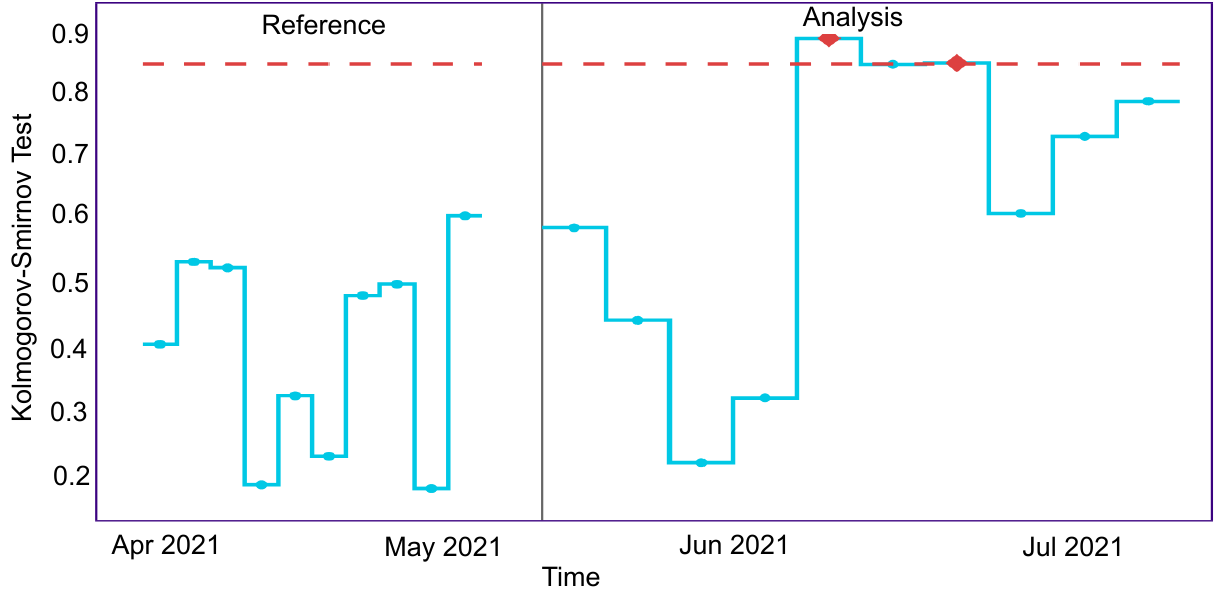}} \newline
    \subfloat[Predictive accuracy vs. KST Test of \texttt{CO2} variable]{\label{fig:nannyml_co2_uc1_predict}\includegraphics[width=\columnwidth]{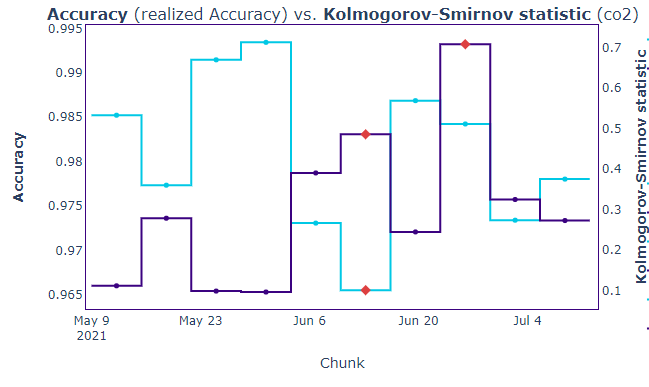}} \newline
    \subfloat[Predictive accuracy vs. KST Test of \texttt{Temperature} variable]{\label{fig:nannyml_temp_uc1_predict}\includegraphics[width=\columnwidth]{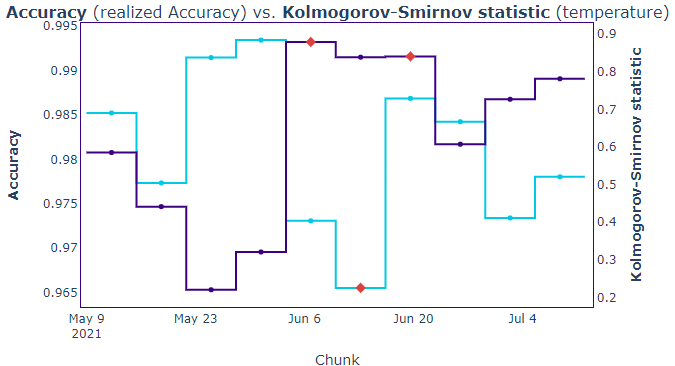}}
    \caption{Reports generated by NannyML for UC1}
	\label{fig:nannyml_results_uc1} 
\end{figure}

The NannyML report facilitates the detection of drift at specific times within the test data. As an example, the Kolmogorov-Smirnov Test reveals drift in 20\% of the chunks for each input variable. Figure~\ref{fig:nannyml_co2_uc1} displays the \texttt{CO2} variable's test outcomes, with alarms activated for June and July 2021. Similarly, Figure~\ref{fig:nannyml_temp_uc1} illustrates the \texttt{temperature} variable's test results, also triggering alarms during the same period. These findings prompt an investigation into any changes in the prediction model's performance. NannyML offers a feature to assess the model's accuracy. Figure~\ref{fig:nannyml_co2_uc1_predict} compares the model's accuracy (in light blue) with the CO2 variable's test results (in purple).

An accuracy drop below 97\% triggers an alarm for reduced predictive accuracy. During the June 9 to June 16 period, model performance falls below this threshold, coinciding with a data drift in the \texttt{CO2} variable detected by the Kolmogorov-Smirnov Test, suggesting an impact on accuracy. An additional alarm, occurring two chunks later, does not affect accuracy. Conversely, the Kolmogorov-Smirnov Test for \texttt{temperature} (cf. Figure~\ref{fig:nannyml_temp_uc1_predict}) does not align with the accuracy drop, although drifts are noted in the chunks immediately before and after the accuracy alarm. This implies that \texttt{temperature} changes might indirectly affect accuracy, despite the lack of exact overlap with the alarms.
In summary, correlating the model's accuracy with the Kolmogorov-Smirnov Test results from NannyML reveals an indirect impact of the temperature variable on the prediction. In contrast, the CO2 variable shows a direct correlation with the model's reduced accuracy.

\textbf{Alibi-Detect:} Table~\ref{tab:uc1_results} presents the outcomes of three Alibi-Detect drift detection methods. The Spot-The-Difference Test identifies drift within the entire dataset rather than isolating specific variables, hence it indicates a data drift but does not distinguish the impact on the two variables. Conversely, both the Kolmogorov-Smirnov and Cramér-von-Mises Tests detect drift in the input variable, yet their results are inconsistent when applied to the target variable. Moreover, the table indicates that only the Kolmogorov-Smirnov Test accurately detects data drifts across all variables (Group One). The Cramér-von-Mises Test incorrectly flags the target variable, classifying it in Group Two. The Spot-The-Difference Test's outcome is uncategorized (N/A) since it does not provide variable-specific results. However, it correctly identifies a Dataset Shift, confirming that $P_{tr}(y, x) \neq P_{tst}(y, x)$.

\subsection{Results of UC2}
%
Table~\ref{tab:uc2_results} summarizes the performance of three open-source detection tools in UC2. 

\textbf{Evidently AI:} Notably, Evidently AI identifies data drifts in both the target and input variables using nine out of twelve methods tested (Group 1). Group 2 contains only the Energy Distance method, which measures the distance between two random vectors. In this method, a calculated distance below 0.1 indicates no detected drift in the target variable's training and test distributions. The Mann-Whitney U-Rank Test, Group 3's sole method, identifies no data drift in the input variable. With a p-value of one, it suggests a 100\% probability of a Type I error, far exceeding the 5\% significance level, leading to the acceptance of the null hypothesis of identical cumulative distribution functions. Finally, Group 4, featuring only the Kullback-Leibler Divergence, does not detect a data drift in any variable. The divergence value returned by Evidently AI is below the threshold of 0.1.
\begin{table*}[htbp]
\centering
\caption{UC2 results using three open-source tools, indicating correct decisions with \checkmark for all variables.}
\label{tab:uc2_results}
\resizebox{\textwidth}{!}{%
\begin{NiceTabular}{@{}lccccccccccccccc@{}}
\toprule
\textbf{Tool} &
  \multicolumn{5}{c}{\textbf{Evidently AI}} &
  \multicolumn{5}{c}{\textbf{NannyML}} &
  \multicolumn{5}{c}{\textbf{Alibi-Detect}} \\ \midrule
Variable &
  \multicolumn{2}{c}{consumption} &
  \multicolumn{2}{c}{temp\_outside} &
   &
  \multicolumn{2}{c}{consumption} &
  \multicolumn{2}{c}{temp\_outside} &
   &
  \multicolumn{2}{c}{consumption} &
  \multicolumn{2}{c}{temp\_outside} &
   \\ \cmidrule(r){1-5} \cmidrule(lr){7-10} \cmidrule(lr){12-15}
Methods &
  \multicolumn{1}{l}{D-Value} &
  \multicolumn{1}{l}{Drift?} &
  \multicolumn{1}{l}{D-Value} &
  \multicolumn{1}{l}{Drift?} &
  \multicolumn{1}{l}{Group} &
  \multicolumn{1}{l}{D-Value} &
  \multicolumn{1}{l}{Drift?} &
  \multicolumn{1}{l}{D-Value} &
  \multicolumn{1}{l}{Drift?} &
  \multicolumn{1}{l}{Group} &
  \multicolumn{1}{l}{D-Value} &
  \multicolumn{1}{l}{Drift?} &
  \multicolumn{1}{l}{D-Value} &
  \multicolumn{1}{l}{Drift?} &
  \multicolumn{1}{l}{Group} \\ \midrule
Kolmogorov-Smirnov Test &
  $2,51e^{-35}$ &
  \checkmark &
  $3,37e^{-48}$ &
  \checkmark &
  1 &
  0,44 &
  \ding{55} (0\%) &
  0,416 &
  \ding{55} (0\%) &
  4 &
  $2,51e^{-35}$ &
  \checkmark &
  0,0 &
  \checkmark &
  1 \\
Wasserstein Distance &
  0,26 &
  \checkmark &
  0,19 &
  \checkmark &
  1 &
  0,13 &
  \ding{55} (20\%) &
  7,31 &
  \ding{55} (0\%) &
  4 &
  \Block[fill=gray!15]{1-5}{} &
   &
   &
   &
   \\
Kullback-Leibler Divergence &
  0,07 &
  \ding{55} &
  0,08 &
  \ding{55} &
  4 &
  \Block[fill=gray!15]{1-10}{} &
   &
   &
   &
   &
   &
   &
   &
   &
   \\
Population Stability Index &
  0,24 &
  \checkmark &
  0,22 &
  \checkmark &
  1 &
  \Block[fill=gray!15]{1-10}{} &
   &
   &
   &
   &
   &
   &
   &
   &
   \\
Jensen-Shannon Distance &
  0,15 &
  \checkmark &
  0,15 &
  \checkmark &
  1 &
  0,51 &
  \checkmark (100\%) &
  0,48 &
  \checkmark (100\%) &
  1 &
  \Block[fill=gray!15]{1-5}{} &
   &
   &
   &
   \\
Anderson Darling Test &
  0,00 &
  \checkmark &
  0,00 &
  \checkmark &
  1 &
  \Block[fill=gray!15]{1-10}{} &
   &
   &
   &
   &
   &
   &
   &
   &
   \\
Cramér-von-Mises Test &
  $6,43e^{-9}$ &
  \checkmark &
  $1,98e^{-10}$ &
  \checkmark &
  1 &
  \Block[fill=gray!15]{1-5}{} &
   &
   &
   &
   &
  $6,43e^{-9}$ &
  \checkmark &
  $1,98e^{-10}$ &
  \checkmark &
  1 \\
Hellinger Distance &
  0,15 &
  \checkmark &
  0,15 &
  \checkmark &
  1 &
  0,47 &
  \checkmark (100\%) &
  0,44 &
  \checkmark (100\%) &
  1 &
  \Block[fill=gray!15]{1-5}{} &
   &
   &
   &
   \\
Mann-Whitney U-Rank Test &
  $1,26e^{-16}$ &
  \checkmark &
  1.00 &
  \ding{55} &
  3 &
  \Block[fill=gray!15]{1-10}{} &
   &
   &
   &
   &
   &
   &
   &
   &
   \\
Energy Distance &
  0,06 &
  \ding{55} &
  0,42 &
  \checkmark &
  2 &
 \Block[fill=gray!15]{1-10}{}  &
   &
   &
   &
   &
   &
   &
   &
   &
   \\
Epps-Singleton Test &
  $2,82e^{-223}$ &
  \checkmark &
  $4,22e^{-234}$ &
  \checkmark &
  1 &
 \Block[fill=gray!15]{1-10}{}  &
   &
   &
   &
   &
   &
   &
   &
   &
   \\
T-Test &
  $8,84e^{-50}$ &
  \checkmark &
  $2,60e^{-16}$ &
  \checkmark &
  1 &
 \Block[fill=gray!15]{1-10}{}  &
   &
   &
   &
   &
   &
   &
   &
   &
   \\
Spot-The-Difference Test &
  \multicolumn{1}{l}{\Block[fill=gray!15]{1-5}{}} &
  \multicolumn{1}{l}{} &
  \multicolumn{1}{l}{} &
  \multicolumn{1}{l}{} &
  \multicolumn{1}{l}{} &
  \Block[fill=gray!15]{1-5}{} &
   &
   &
   &
   &
  0,07 &
  NA &
  0.07 &
  NA &
  NA \\ \bottomrule
\end{NiceTabular}%
}
\end{table*}

\begin{figure}[htbp]
	\centering
	\subfloat[Consumption variable]{\label{fig:evidently_cons}\includegraphics[width=0.96\columnwidth]{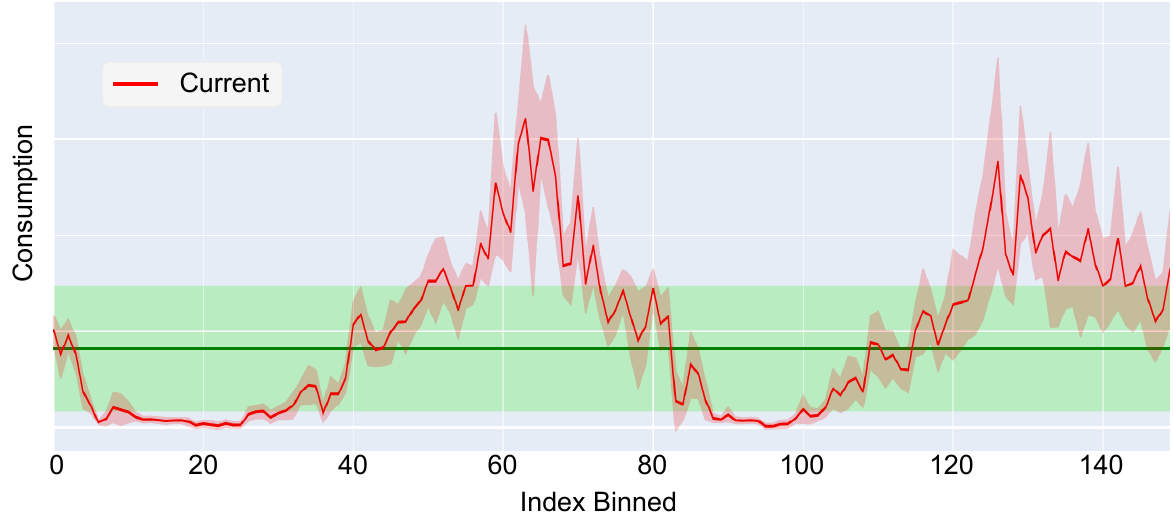}} \newline
	\subfloat[Distributions of Consumption variable]{\label{fig:evidently_dist}\includegraphics[width=\columnwidth]{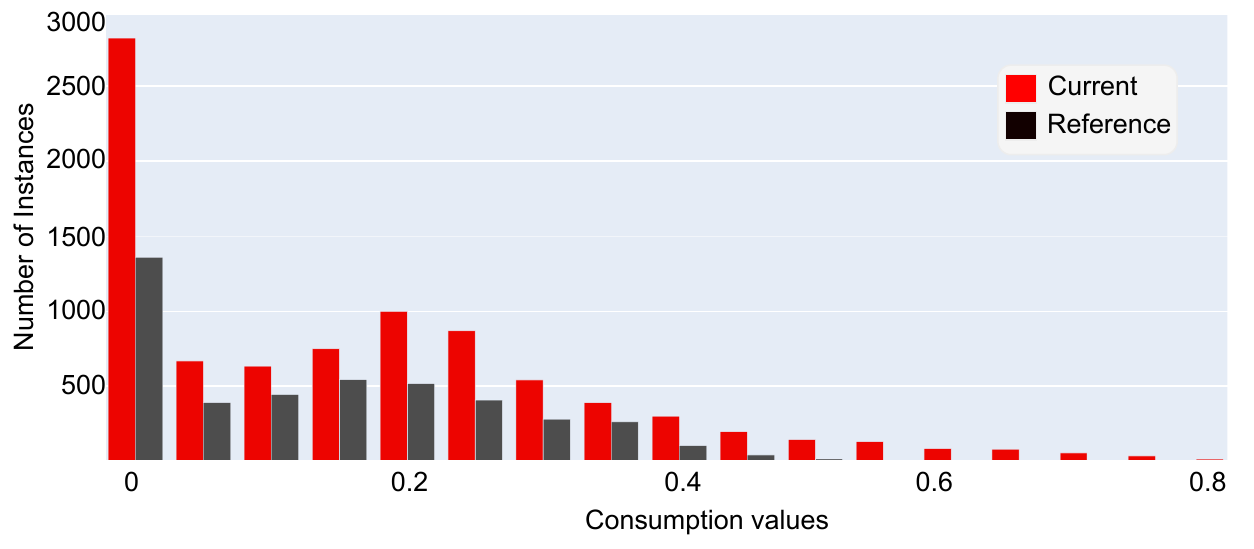}}
	\caption{Reports generated by Evidently AI for UC2}
	\label{fig:evidently_results} 
\end{figure}

Evidently AI creates a detailed results report, exemplified and analyzed using the Kolmogorov-Smirnov Test in Figure~\ref{fig:evidently_results}. It begins with an overview that examines the data drift across the entire dataset, followed by detailed views that explore drifts in individual variables. For example, Figure~\ref{fig:evidently_cons} shows the values of the consumption variable (i.e., target) versus the time variable. The green line indicates the training data's average, while the shaded green area denotes its standard deviation. The graph reveals four instances where consumption values fall outside the standard deviation, attributed to seasonal fluctuations in energy use. However, the graph does not detail the Kolmogorov-Smirnov Test calculations. Figure~\ref{fig:evidently_dist} compares the distributions between the training (reference) and test (current) data. The Kolmogorov-Smirnov Test uses these distributions to evaluate the differences in distribution functions. Theoretically, if both distributions were identical, the test data bars on the graph, which are double the quantity of the training data, would be twice the length at every x-value. However, Figure~\ref{fig:evidently_dist} illustrates that this condition does not hold for multiple x-values.

The Evidently AI's report indicates that all examined variables show drift, with the \quotes{Share of Drifted Columns} at 100\%, surpassing the threshold of 50\%. The \quotes{Data Drift Summary} details each variable, including data type and distribution in training and test datasets, and the drift value and method used. Notably, the drift value is zero, corresponding to the Kolmogorov-Smirnov Test's p-value. This suggests complete confidence in rejecting the null hypothesis of equal distribution functions, accepting the alternative that they are unequal, with no risk of a Type I error.

\begin{figure}[htbp]
	\centering
    \subfloat[Wasserstein Distance for the consumption variable]{\label{fig:nannyml_cons}\includegraphics[width=\columnwidth]{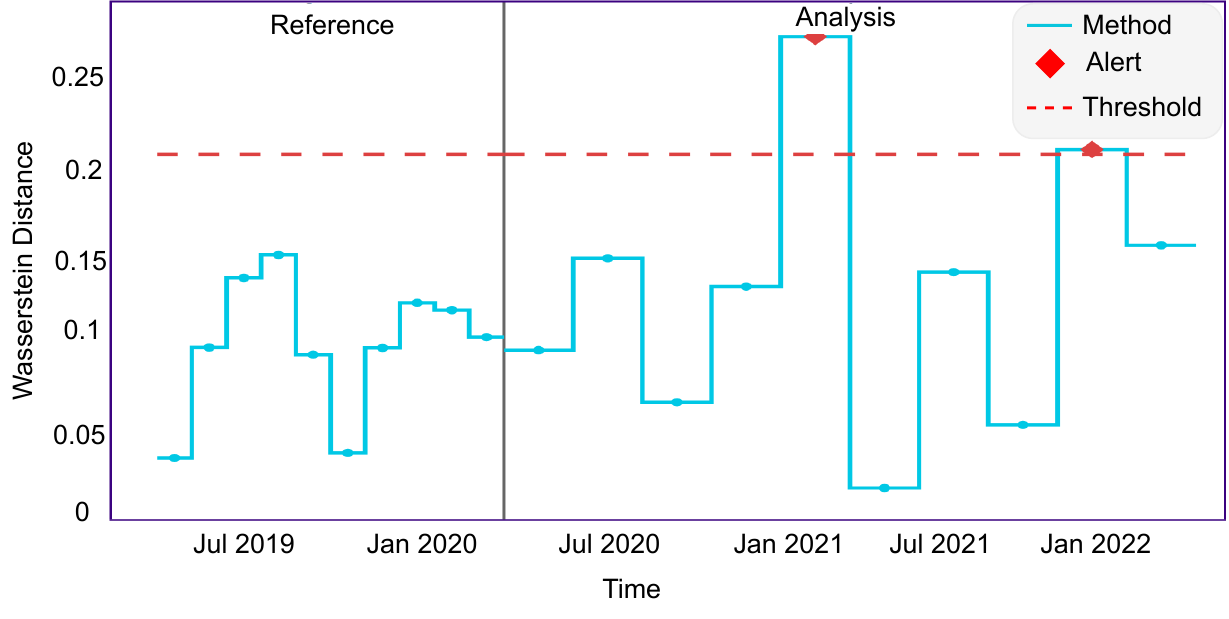}} \newline
    \subfloat[NannyNL alerts per chunk for the consumption variable]{\label{fig:nannyml_chunks}\includegraphics[width=\columnwidth]{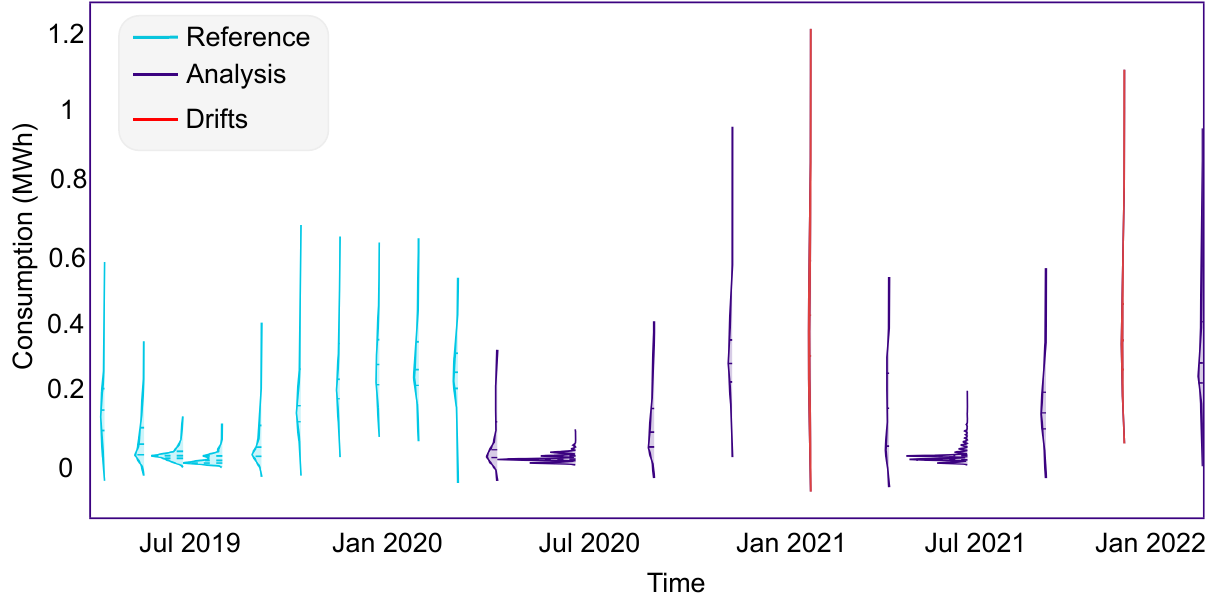}}
	\caption{Reports generated by NannyML for UC2}
	\label{fig:nannyml_figures} 
\end{figure}

\textbf{NannyML: } NannyML identifies drifts by analyzing data chunks, and a drift is considered recognized if it is detected in at least 50\% of these chunks. Table~\ref{tab:uc2_results} indicates that half of the four evaluated methods detect drifts in both the target and input variables. However, the Kolmogorov-Smirnov Test fails to detect drift in any variables, while the Wasserstein Distance does not register drift in any variables, despite noting a drift in the target variable in 20\% of the chunks. NannyML provides the exact timing of drifts by returning the start and end dates of affected chunks through its interface. NannyML produces a result report featuring two key figures, as shown in Figure~\ref{fig:nannyml_figures}. The first illustrates the method's test statistic over time, while the second displays identified data drifts across data chunks. 

As an example, Figure~\ref{fig:nannyml_cons} shows the Wasserstein Distance for the \texttt{consumption} variable. Such a measure detected a drift in 20\% of the chunks (cf. Table~\ref{tab:uc2_results}). The red dashed line in the figure indicates a predetermined threshold. An alarm is triggered when a Wasserstein Distance value surpasses this threshold. The figure's left portion depicts this distance for the training data (Reference) and the right portion for the test data (Analysis). The figure shows two instances of detected shifts by the Wasserstein Distance, both occurring in winter. This aligns with observations from Section~\ref{sec:cases} that suggest significant fluctuations in the test data, especially during the winter months. Pinpointing the precise timing of shifts within the dataset is a notable aspect of this analysis.

Figure~\ref{fig:nannyml_chunks} depicts the second graph in the NannyML report illustrates data distribution across chunks, highlighting detected shifts. This ridgeline plot displays rotated density distributions for each chunk, which may overlap. Inside each plot, three lines represent the first, second, and third quartiles. The plot's vertical extent shows the range from minimum to maximum values. The figure marks chunks with detected data drift in red. In the test (Analysis) data, two chunks are highlighted in red, signaling significant dispersion as evidenced by substantial differences in the minimum and maximum values and obscured quartile lines in the Joyplots. These red chunks, exhibiting the highest energy values, correspond with the largest spreads and values, leading to increased Wasserstein Distances and threshold breaches. Notably, these chunks both fall within winter months, aligning with Section~\ref{sec:cases}'s observation of pronounced variability during this season.

\textbf{Alibi-Detect:} Table~\ref{tab:uc2_results} presents the detection capabilities of Alibi-Detect, with two out of three methods identifying a data drift in both the target and input variables. As listed in the table, the Spot-The-Difference Test fails to detect a data drift. This test was unable to calculate a test variable for each variable, leading to a single drift value computed for the entire dataset, reflected identically in both variable columns. The Kolmogorov-Smirnov and Cramér-von-Mises Tests accurately identify data drifts in both variables (Group one). The Spot-The-Difference Test's outcome cannot be classified within these groups (effectively, NA) since it provides no variable-specific results. Ultimately, it does not detect a data drift. Alibi-Detect does not provide visualizations or reports; hence, no results report is included.

\subsection{Non-Functional Comparisons}
\begin{figure}[htbp]
	\centering
    \subfloat[UC1]{\label{fig:time_uc1}\includegraphics[width=\columnwidth]{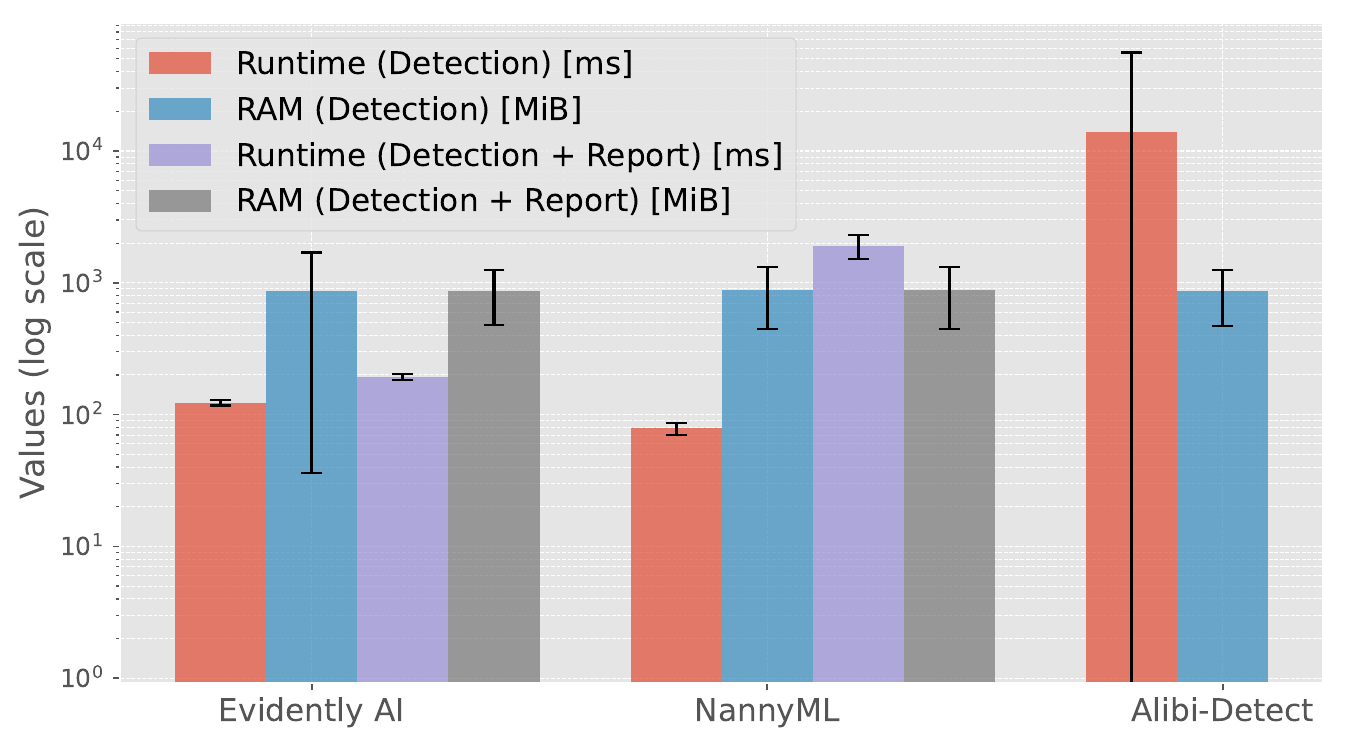}} \newline
    \subfloat[UC2]{\label{fig:time_uc2}\includegraphics[width=\columnwidth]{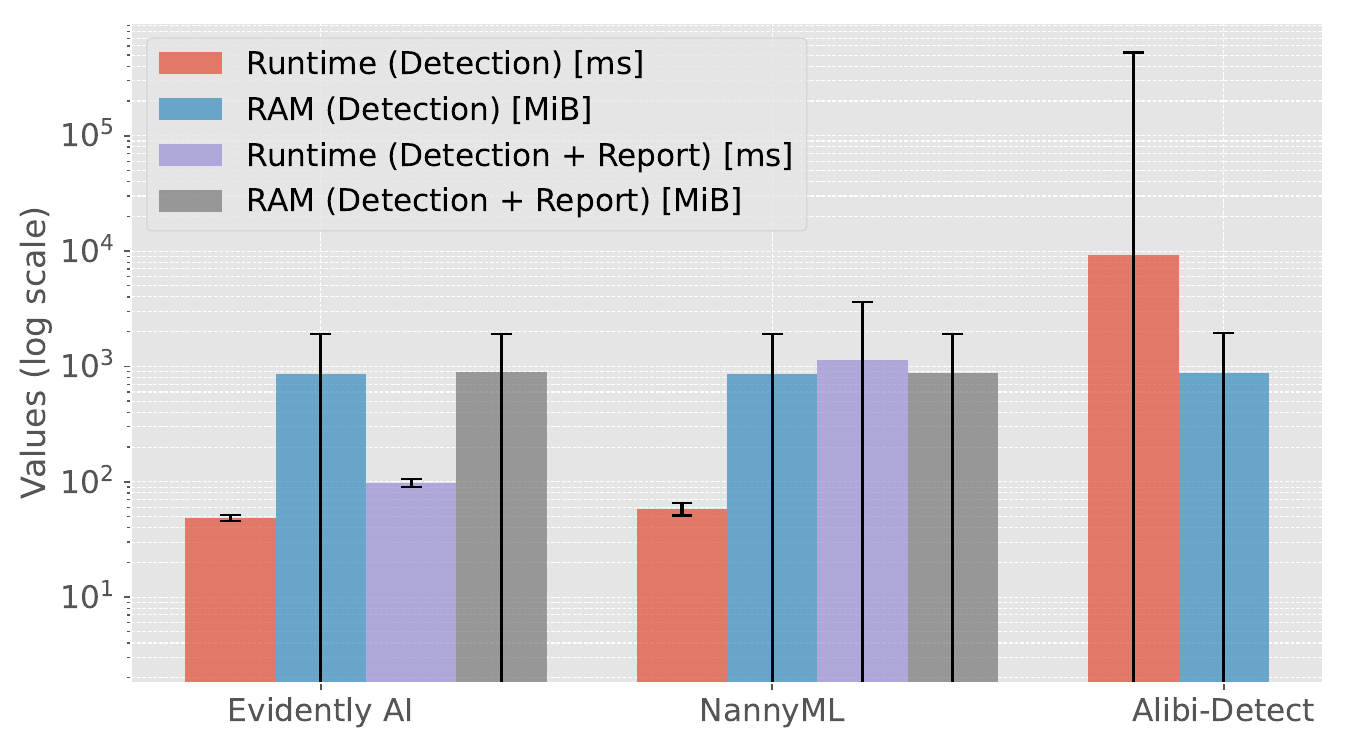}}
	\caption{Runtime and RAM consumption of the compared tools}
	\label{fig:time_figures} 
\end{figure}

%
\textbf{Runtime \& RAM Consumption:} Figure~\ref{fig:time_figures} compares the runtime and RAM usage of three tools, both with and without report generation. Alibi-Detect, which does not produce reports, has metrics only for the latter case. The presented values represent the means from five trials for each tool. For UC1, Figure~\ref{fig:time_uc1} depicts that Alibi-Detect's runtime is substantially longer than the other tools, exceeding NannyML's by approximately 17,840\% and Evidently AI's by about 11,366\%. Conversely, Evidently AI's runtime is approximately 57\% greater than NannyML's. In terms of RAM usage, NannyML requires approximately 1.5\% more RAM than Evidently AI, while Alibi-Detect is the most efficient, using about 0.5\% less RAM than Evidently AI and around 2\% less than NannyML. For Evidently AI, running the tool for detection along with report generation increases the runtime by about 57\%, while the RAM usage slightly decreases by approximately 0.2\%. In the case of NannyML, the runtime sees a substantial increase of approximately 2330\% when adding report generation, with virtually no change in RAM usage.

For UC2, Figure~\ref{fig:time_uc2} shows that Evidently AI and NannyML have comparable detection runtimes at 48.26 ms and 57.92 ms, respectively, with NannyML being approximately 20\% slower. Alibi-Detect's runtime is significantly longer, at 9219.69 ms, making it roughly 19,000\% slower than Evidently AI. The reason for this is the Spot-The-Difference Test, as it trains a classifier that differentiates between test and training data. Moreover, Alibi-Detect's RAM usage for detection is slightly higher, at about 2.6\% more than Evidently AI's and 2.7\% more than NannyML's, while Evidently AI's RAM consumption is marginally higher, by approximately 0.3\%, compared to NannyML. The figure shows that Alibi-Detect has a much larger variability in both runtime and RAM usage during detection compared to Evidently AI and NannyML. Meanwhile, Evidently AI and NannyML show more consistency in runtime but have a considerable variance in RAM usage, suggesting variability in their memory consumption. When comparing the runtime for detection alone to detection with report generation, Evidently AI shows an increase of over 103\%, while NannyML exhibits a dramatic increase, almost 1876\%. For RAM usage, Evidently AI sees a slight increase of about 2.6\% when generating reports, whereas NannyML shows a marginal decrease of approximately 0.3\% in RAM consumption in the report generation scenario.

The comparisons suggest that Alibi-Detect could be less suitable for time-sensitive applications given its long runtime, although it is slightly more RAM efficient in UC1. Evidently AI and NannyML are more time-efficient but show considerable variance in RAM usage, which could be a factor in environments with limited memory resources. When generating reports, Evidently AI offers a more efficient and stable performance in both use cases, with significantly shorter runtimes and less variability in the measurements. The large increase in runtime for NannyML during report generation also indicates that users should be prepared for significantly longer processing times in such scenarios. 

\textbf{Tool Selection:} While the tools are capable of detecting drifts in both use cases, the effectiveness of the methods varies, requiring careful selection of both the tool and the methods it offers. For UC2, Evidently AI detects data drift with nine methods, surpassing the six methods effective in UC1. Five methods consistently identify drift in both use cases: Kolmogorov-Smirnov Test, Wasserstein Distance, Population Stability Index, Jensen-Shannon Distance, and Hellinger Distance. Energy Distance is only accurate in UC2. Evidently AI's report indicates the presence of data drift and identifies the affected variables, but the exact timing of drift is not specified through its interface. 

In both UC1 and UC2, NannyML identifies drift correctly with two methods, specifically the Jensen-Shannon and Hellinger Distances. The Wasserstein Distance also detects drift in certain chunks for both cases, with the Kolmogorov-Smirnov Test doing the same in UC1. The report, together with the classification model's accuracy from UC1, shows the impact of input variables on model performance. NannyML's report details drift detection in variables and the specific chunks of the test data where it occurs. Start and end dates for these chunks are accessible through the NannyML interface. Alibi-Detect's Kolmogorov-Smirnov Test reliably detects data drift in both use cases, while the Cramér-von-Mises Test succeeds only in UC2 but not in UC1. The Spot-The-Difference Test inconsistently identifies drift, only doing so for the entire dataset in UC1. Alibi-Detect does not provide the exact timing of shift occurrences in test data and lacks in-depth result visualization. 

\textbf{Methods Implementation:} There are significant differences in how the same methods are implemented across various tools. For instance, the Kolmogorov-Smirnov Test, implemented in all tools, shows identical drift values in both Evidently AI and Alibi-Detect. Their implementations align, utilizing the \texttt{scipy.stats} library and the \texttt{ks\_2samp} function for the two-sample test. NannyML processes training data based on its size, removing missing values before calculations. For datasets with under 10,000 records, it performs an exact Kolmogorov-Smirnov Test, comparing the full training set to the test data and yielding a p-value. For datasets larger than 10,000 records, NannyML executes an approximate Kolmogorov-Smirnov Test, dividing both training and test data into quartiles and calculating their cumulative relative frequency distributions. The maximum discrepancy between these distributions is then assessed.

\textbf{Integrability:} Evidently AI can be installed as a PyPI package using pip or conda. While visualizations work directly in Jupyter Notebooks. The reports are also compatible with cloud notebooks including Google Colab (officially supported), Kaggle Notebook, Jupyter Lab, Deepnote, and Databricks. Integration with MLflow, Airflow, Metaflow, Prefect, FastAPI, and PostgreSQL allows for incorporation into machine learning project workflows. Additionally, Evidently AI supports email notifications via AWS SES. NannyML can be installed via pip or conda, or as a Docker container using a pre-built Docker image. It may require additional binaries for the LightGBM library, based on the operating system. NannyML integrates with PostgreSQL and Grafana for ML project development and monitoring. Finally, Alibi-Detect is accessible as a PyPI package installable through pip or conda (via conda-forge). It requires the selection of a backend during installation, offering choices like TensorFlow, Pytorch, KeOps, and Prophet. It also supports ML tools such as Seldon Core, KServe, and Ray for enhanced functionality.

\textbf{Flexibility:} Evidently AI provides detection methods for both tabular and text data, analyzing aspects like text length, out-of-vocabulary (OOV) word similarity, and embeddings. It can identify data drifts at the level of individual variables or across the entire dataset for numerical, categorical, and binary data. NannyML specializes in tabular datasets compatible with Pandas DataFrames, detecting drifts in individual variables or the whole dataset, with support limited to numerical and categorical data. Finally, Alibi-Detect offers a versatile range of detection methods for tabular, image, time series, and text data. It operates in online and offline modes, capable of detecting drifts in both categorical and numerical variables for individual features or complete datasets, tailored to the type of data.

\textbf{User-friendliness:} Evidently AI stands out for its extensive, user-friendly documentation complemented by practical examples in Jupyter and Google Colab Notebooks, video demos, and presets. Visual reports further simplify result interpretation. However, the platform lacks clear guidance on extracting method-specific drift values and does not aid in selecting training and test data, assuming users pre-identify data drifts. Some reports are also missing axis labels, complicating the result analysis. In contrast, NannyML differentiates itself with detailed scientific documentation on detection methods and provides example notebooks for easier onboarding. It offers easy-to-understand visualizations, though these require individual generation. Like Evidently AI, it does not guide users in selecting training and test datasets. 

Alibi-Detect's user experience suffers from initial backend choices affecting runtime or available detection methods. Each method requires separate implementation and import, and the platform only offers a limited set of univariate methods for numerical columns without PyTorch, TensorFlow, or a binary classification model. Implementation challenges such as array type conversion arose, and despite claims, the Spot-The-Difference Test could not detect drifts for individual variables in practice. Alibi-Detect also lacks assistance in selecting training and test data.

\subsection{Lessons Learned}

The analysis indicates that Evidently AI excels in detecting general data drifts, particularly in scenarios where predicted model values are unavailable. Conversely, NannyML is exceptionally adept at pinpointing the precise moment of a data drift and correlating it with the accuracy of predictions. Evidently AI also distinguishes itself with its smooth integration into ML software, earning high marks for user-friendliness. Its compatibility with MLflow and Grafana positions it as the preferred choice for seamless incorporation into ML pipelines and enhancing the MLOps workflow. Despite these strengths, Evidently AI does not currently offer support for image data analysis. In this domain, Alibi-Detect emerges as the sole tool providing this capability, highlighting its unique position in the toolset for ML practitioners.

Based on the findings, we recommend a streamlined three-step process to effectively identify and manage data drift in time series data. Initially, one should assess the impact of potential drift on the model's predictive accuracy. This can often be performed within existing ML monitoring pipelines and may not require additional tools, though NannyML is available for this purpose. A significant decrease in model accuracy can indicate data drift, necessitating further examination in the subsequent step. Conversely, if the accuracy remains stable, data drift is unlikely to be influencing the model's predictions.

Should the first step suggest a drift, the second step involves confirming its presence using Evidently AI, which excels in analyzing both the entire dataset and individual variables for signs of drift. Employing statistical measures such as the Kolmogorov-Smirnov Test, Wasserstein Distance, Population Stability Index, Jensen-Shannon Distance, and Hellinger Distance will provide a comprehensive drift analysis. Following confirmation, NannyML comes into play for the third step—determining the exact timing of the drift and its effect on prediction accuracy. NannyML's precision with methods like the Kolmogorov-Smirnov Test and Wasserstein Distance is instrumental in pinpointing when and where the drift occurred. This final analysis allows for a full correlation with the initial accuracy assessment, thereby completing the drift detection process. It is essential to use both Evidently AI and NannyML to navigate through these steps and achieve a thorough detection and understanding of data drift.
%
%
\section{Conclusion \& Future Work}\label{sec:conclusion}
%
In this study, we conducted a comparative analysis of three open-source tools—Evidently AI, NannyML, and Alibi-Detect—focused on detecting data drifts utilizing datasets from smart building applications. The tools were assessed based on a spectrum of functional and non-functional criteria using real-world time series data. Our findings highlighted that Evidently AI excels at identifying broad data drifts. NannyML, on the other hand, stands out for its precision in pinpointing the specific timing of drifts and evaluating their impact on the accuracy of predictions. The practical deployment of these tools hinges on their functional suitability, and in this regard, Evidently AI emerged as the preferred option. Its rapid computational capabilities, along with its adeptness at generating insightful reports and seamless integration with existing ML tools.

This research focused exclusively on univariate detection methods, leaving the exploration of multivariate approaches for future investigation. By comparing and analyzing multivariate detection methods, we can expect to develop more sophisticated and robust detection mechanisms. Additionally, there is a compelling need to delve into the detection of data drifts within image data. Given its intricate structures and patterns, image data poses unique challenges that are crucial to address, especially considering its significance in domains such as computer vision and medical imaging.

\bibliographystyle{plain}
\bibliography{main}

\end{document}